\documentclass[journal]{IEEEtran}
\ifCLASSINFOpdf
\else
\fi
\usepackage{amsmath}
\usepackage[nospace,noadjust]{cite}
\usepackage{amsfonts}
\usepackage{caption}
\usepackage[usenames,dvipsnames,table]{xcolor}
\usepackage{multirow}
\usepackage{booktabs}
\usepackage{pgfplots}
\usepackage{smartdiagram}
\usepackage{metalogo}
\usepackage{balance}
\usepackage[caption=false]{subfig}
\usepgfplotslibrary{units}
\usepgfplotslibrary{groupplots}
\usetikzlibrary{patterns}
\usepackage{tikz}
\usepackage{pgfplots}
\usetikzlibrary{arrows,shapes,backgrounds,fit}
\usepgfplotslibrary{colorbrewer}
\usetikzlibrary{calc,positioning,shapes,decorations.pathreplacing,snakes}
\usepackage{graphicx}
\DeclareCaptionLabelSeparator{periodspace}{.\quad}
\usepackage[nolist]{acronym}

\newcommand{\squeezeup}{\vspace{-3.5mm}}

\usepackage{hyperref}
\pgfplotsset{compat=1.14}

\begin{document}
%
\title{
 Deep Learning Models for Wireless Signal Classification with Distributed Low-Cost Spectrum Sensors
}
%
%
%
 \author{Sreeraj~Rajendran,~\IEEEmembership{Student Member,~IEEE,}
        Wannes~Meert,~\IEEEmembership{Member,~IEEE}
        Domenico Giustiniano,~\IEEEmembership{Senior Member,~IEEE,}
        Vincent Lenders,~\IEEEmembership{Member,~IEEE}
        and~Sofie~Pollin,~\IEEEmembership{Senior Member,~IEEE.}

}

\maketitle

\begin{abstract}
This paper looks into the modulation classification problem for a distributed wireless spectrum sensing network. First, a new data-driven model for \ac{amc} based on long short term memory (LSTM) is proposed. The model learns from the time domain amplitude and phase information of the modulation schemes present in the training data without requiring expert features like higher order cyclic moments. Analyses show that the proposed model yields an average classification accuracy of close to 90\% at varying SNR conditions ranging from 0dB to 20dB. Further, we explore the utility of this LSTM model for a variable symbol rate scenario. We show that a LSTM based model can learn good representations of variable length time domain sequences, which is useful in classifying modulation signals with different symbol rates. The achieved accuracy of 75\% on an input sample length of 64 for which it was not trained, substantiates the representation power of the model. To reduce the data communication overhead from distributed sensors, the feasibility of classification using averaged magnitude spectrum data and on-line classification on the low-cost spectrum sensors are studied. Furthermore, quantized realizations of the proposed models are analyzed for deployment on sensors with low processing power.

\end{abstract}

\begin{IEEEkeywords}
Deep learning, Modulation classification, LSTM, CNN, Spectrum sensing.
\end{IEEEkeywords}

%
\IEEEpeerreviewmaketitle

\section{Introduction}
%
%
%
%

 

\begin{acronym}[HBCI]
%
%
%
%
%

\acro{3gpp}[3GPP]{3\textsuperscript{rd} Generation Partnership Program}
\acro{cnn}[CNN]{Convolutional Neural Network}
\acro{fbmc}[FBMC]{Filter Bank Multicarrier}
\acro{phy}[PHY]{physical layer}
\acro{pu}[PU]{Primary User}
\acro{rat}[RAT]{Radio Access Technology}
\acro{rfnoc}[RFNoC]{RF Network on Chip}
\acro{sdr}[SDR]{Software Defined Radio}
\acro{su}[SU]{Secondary User}
\acro{toa}[TOA]{Time of Arrival}
\acro{tdoa}[TDOA]{Time Difference of Arrival}
\acro{usrp}[USRP]{Universal Software Radio Peripheral}
\acro{amc}[AMC]{Automatic Modulation Classification}
\acro{lstm}[LSTM]{Long Short Term Memory}
\acro{soa}[SoA]{state-of-the-art}
\acro{fft}[FFT]{Fast Fourier Transform}
\acro{wsn}[WSN]{Wireless Sensor Networks}
\acro{iq}[IQ]{in-phase and quadrature phase}
\acro{snr}[SNR]{signal-to-noise ratio}
\acro{sps}[sps]{samples/symbol}
\acro{awgn}[AWGN]{Additive White Gaussian Noise}
\acro{ofdm}[OFDM]{Orthogonal Frequency Division Multiplexing}
\acro{rnn}[RNN]{Recurrent Neural Networks}
\acro{svm}[SVM]{Support Vector Machines}
\acro{psd}[PSD]{Power Spectral Density}
\end{acronym}

Wireless spectrum monitoring over frequency, time and space is important for a wide range of applications such as spectrum enforcement for regulatory bodies, generating coverage maps for wireless operators, and applications including wireless signal detection and positioning. Continuous spectrum monitoring over a large geographical area is extremely challenging mainly due to the multidisciplinary nature of the solution. The monitoring infrastructure requires proper integration of new disruptive technologies than can flexibly address the variability and cost of the used sensors, large spectrum data management, sensor reliability, security and privacy concerns, which can also target a wide variety of the use cases. Electrosense was designed to address these challenges and support a diverse set of applications \cite{electrosense}. Electrosense is a crowd-sourced spectrum monitoring solution deployed on a large scale using low cost sensors.

One of the main goals of Electrosense is to accomplish automated wireless spectrum anomaly detection, thus enabling efficient spectrum enforcement. Technology classification or specifically Automatic Modulation Classification (\ac{amc}) is an integral part of spectrum enforcement. Such a classifier can help in identifying suspicious transmissions in a particular wireless band. Furthermore, technology classification modules are fundamental for interference detection and wireless environment analysis. Considering the aforementioned large application space this paper looks into two key aspects: Is efficient wireless technology classification achievable on a large scale with low cost sensor networks and limited uplink communication bandwidth? If possible, which are the key classification models suitable for the same.

The number of publications related to \ac{amc} appearing in literature is large \cite{txminer,dof,litsurvey,gardner_unif,zaihe_thesis,cyclo_test} mainly due to the broad range of problems associated with \ac{amc} and huge interest in the problem itself for surveillance applications. \ac{amc} helps a radio system for environment identification, defining policies and taking actions for throughput or reliability improvements. It is also used for applications like transmitter identification, anomaly detection and localization of interference \cite{txminer,dof}. 

Various approaches for modulation classification discussed in literature can be brought down to two categories \cite{litsurvey}, one being the \emph{decision theoretic} approach and the other the \emph{feature based} approach. In decision-theoretic approaches the modulation classification problem is presented as a multiple hypothesis \mbox{testing problem \cite{litsurvey}}. The maximum likelihood criterion is applied to the received signal directly or after some simple transformations such as averaging. Even though decision-theoretic classifiers are optimal in the sense that they minimize the probability of miss-classifications, practical implementations of such systems suffer from computational complexity as they typically require buffering a large number of samples. These methods are also not robust in the presence of unknown channel conditions and other receiver discrepancies like clock frequency offset. 

Conventional feature-based approaches for \ac{amc} make use of expert features like cyclic moments \cite{gardner_unif}. Spectral correlation functions of various analog and digital modulation schemes covered in \cite{scf_analog} and \cite{scf_digital} respectively are the popularly used features for classification. Detailed analysis of various methods using these cyclostationary features for modulation classification are presented in \cite{zaihe_thesis}. Various statistical tests for detecting the presence of cycles in the $k$th-order cyclic cumulants without assuming any specific distribution on the data are presented in \cite{cyclo_test}. In \cite{scf_nn} authors used a multilayer linear perceptron network over spectral correlation functions for classifying some basic modulation types. Another method makes use of the cyclic prefix \cite{ofdm_classif} to distinguish between multi-carrier and single carrier modulation schemes which is used for \ac{ofdm} signal identification.

All these aforementioned model driven approaches exploit knowledge about the structure of different modulation schemes to define the rules for \ac{amc}. This manual selection of expert features is tedious which makes it difficult to model all channel discrepancies. For instance, it is quite challenging to develop models which are robust to 
fading, pathloss, time shift and sample rate variations. In addition, a distributed collection of \ac{iq} data over frequency, space and time is expensive in terms of transmission bandwidth and storage. Furthermore, most of these algorithms are processor intensive and could not be easily deployed on low-end distributed sensors.

Recently, deep learning has been shown to be effective in various tasks such as image classification, machine translation, automatic speech recognition \cite{deeplearn} and network optimization \cite{cobanets}, thanks to multiple hidden layers with non-linear logistic functions which enable learning higher-level information hidden in the data. A recently proposed deep learning based model for \ac{amc} makes use of a \ac{cnn} based classifier \cite{o2016convolutional}. The \ac{cnn} model operates on the time domain \ac{iq} data and learns different matched filters for various \ac{snr}. However, this model may not be efficient on data with unknown sampling rates and pulse shaping filters which the model has never encountered during the training phase. Also being a fixed input length model, the number of modulated symbols the model can process remains limited across various symbol rates. Furthermore, the training and computational complexity of the model increases with increasing input sample length. In \cite{baseline}, the authors extended the analysis on the effect of \ac{cnn} layer sizes and depths on classification accuracy. They also proposed complex inception modules combining \ac{cnn} and \ac{lstm} modules for improving the classification results. In this paper we show that simple \ac{lstm} models can itself achieve good accuracy, if input data is formatted as amplitude and phase (polar coordinates) instead of \ac{iq} samples (rectangular coordinates).

This paper proposes a \ac{lstm} \cite{Hochreiter:lstm} based deep learning classifier solution, which can learn long term temporal representations, to address the aforementioned issues. The proposed variable input length model can capture sample rate variations without explicit feature extraction. We first train the LSTM model to classify 11 typical modulation types, as also used in \cite{o2016radio}, and show our approach outperforms the \ac{soa}. Being a variable input length model we also show that the model enables efficient classification on variable sample rates and sequence lengths. Even though these deep learning models can provide good classification accuracies on lower input sample lengths, their computational power requirements are still high preventing them from low-end sensor deployment as Electrosense.

The wireless sensing nodes deployed in the Electrosense network consist of a low-cost and bandwidth limited \ac{sdr} interfaced with a small sized embedded platform \cite{electrosense}. \ac{psd} and \ac{iq} pipelines are enabled in the sensor to support various applications producing data in the order of 50-100~Kbps and 50~Mbps respectively. First, the embedded hardware of the sensors is not powerful enough to handle performance intensive \ac{amc} algorithms. Second, transferring \ac{iq} samples to the backend for classification by enabling the \ac{iq} pipeline is not a scalable solution as it is expensive in terms of data transfer and storage. Finally, the sensors are bandwidth limited which prevents them from acquiring wideband signals.

To enable instantiation of the newly proposed \ac{lstm} model for modulation classification in a large distributed network of low cost sensor nodes, we compare various approaches to decrease the implementation cost of the classifier. In the first approach we study the advantages and limitations of classification models for modulation classification on a deployed distributed sensor network with limited bandwidth sensors based on averaged magnitude \ac{fft} data which decreases the communication cost by a factor 1000. Moreover, quantized versions of the proposed models are studied in detail for sensor deployment. These quantized versions can be run on a low cost sensor and do not require the instantiation of the classifier in the cloud. As a result, the sensor should only communicate the decision variable, which further decreases the communication cost. The \emph{code and datasets} for all the deep learning models are \emph{made public} for future research\footnote{\label{noterepo}\url{https://github.com/zeroXzero/modulation_classif}}. The models are also available for use through Electrosense. 

The contribution of this paper is thus threefold. First, we develop a new \ac{lstm} based deep learning solution using time domain amplitude and phase samples which provides \ac{soa} results for high \ac{snr}s on a standard dataset. Second, we explore the use of deep learning models for technology classification task in a distributed sensor network only using averaged magnitude \ac{fft} data. Finally, we explore the model performance by quantizing the deep neural networks for sensor deployment.

The rest of the paper is organized as follows. The classification problem is clearly stated in Section~\ref{problem}. A brief overview of the modulation dataset and the channel models used are presented in Section~\ref{dataset}.  Section~\ref{models} explains the \ac{lstm} model used for classification and the parameters used for training along with other implementation details. Section~\ref{results} details the classification results and discusses the advantages of the proposed model. Low-implementation cost models are discussed in Section~\ref{lmodels}. Conclusions and future work are presented in Section~\ref{conclusion}.

\section{Problem Statement}
\label{problem}
Technology or modulation recognition can be framed as a N-class classification problem in general. A general representation for the received signal is given by
\begin{equation}
\begin{aligned}
r(t)&=s(t)*c(t)+n(t),
\end{aligned}
\label{eq_full}
\end{equation}
where $s(t)$ is the noise free complex baseband envelope of the received signal, $n(t)$ is \ac{awgn} with zero mean and variance $\sigma_n^2$ and $c(t)$ is the time varying impulse response of the transmitted wireless channel. The basic aim of any modulation classifier is to give out $P(s(t)\in N_i| r(t))$ with $r(t)$ as the only signal for reference and $N_i$ represents the $i$th class. The received signal $r(t)$ is commonly represented in \ac{iq} format due to its flexibility and simplicity for mathematical operations and hardware design. The in-phase and quadrature components are expressed as $I = A cos(\phi)$ and $Q = A sin(\phi)$, where $A$ and $\phi$ are the instantaneous amplitude and phase of the received signal $r(t)$.

The RadioML and modified RadioML datasets used for testing the proposed model, presented in the next section of this paper, follow the signal representation as given in equation~\ref{eq_full}. These datasets make a practical assumption that the sensor's sampling rate is high enough to receive the full-bandwidth signal of interest at the receiver end as $r(t)$. The datasets also take into account complex receiver imperfections which are explained in detail in Section~\ref{dataset}. The \textit{samples per symbol} parameter used in the tables~\ref{table_rml_dataset} and \ref{table_modrml_dataset} specify the number of samples representing each modulated symbol which is a modulation characteristic. Similarly \textit{sample length} parameter specifies the number of received signal samples used for classification.

\section{Modulation Datasets}
\label{dataset}
\begin{table}[!t]
\begin{center}
\begin{tabular}{|l|l|}
	\hline
    Modulations     & 8PSK, AM-DSB, AM-SSB, BPSK,\\
    								  & CPFSK, GFSK, PAM4, QAM16,\\
    								  & QAM64, QPSK, WBFM\\
    \hline
  	Samples per symbol &   4 \\
  	\hline	
    Sample length &   128 \\
  	\hline
    SNR Range &  -20dB to +20dB \\
  	\hline
   	Number of training samples &   82500 vectors\\
  	\hline
  	Number of test samples &  82500 vectors\\
    \hline
\end{tabular}
\end{center}
\caption{RadioML2016.10a dataset parameters.}
\label{table_rml_dataset}
\end{table}

\begin{table}[!t]
\begin{center}
\begin{tabular}{|l|l|}
	\hline
    Modulations     & 8PSK, AM-DSB, AM-SSB, BPSK,\\
    								  & CPFSK, GFSK, PAM4, QAM16,\\
    								  & QAM64, QPSK, WBFM\\
    \hline
  	Samples per symbol &   4 and 8 sps\\
  	\hline	
    Sample length &   128 to 512\\
  	\hline
    SNR Range &  -20dB to +20dB \\
  	\hline
  	Number of training samples &   165000 vectors\\
  	\hline
  	Number of test samples &  165000 vectors\\
    \hline
\end{tabular}
\end{center}
\caption{Modified complex RadioML dataset parameters.}
\label{table_modrml_dataset}
\end{table}

A publicly available dataset used for evaluating the performance of the proposed model is detailed in this section. The standard dataset is also extended to evaluate the sample rate dependence of the proposed model.

\subsection{RadioML dataset}
A standard modulation dataset presented in \cite{o2016radio} is used as the baseline for training and evaluating the performance of the proposed classifier. The used RadioML2016.10a dataset is a synthetically generated dataset using GNU Radio \cite{gnuradio_web} with commercially used modulation parameters. This dataset also includes a number of realistic channel imperfections such as channel frequency offset, sample rate offset, additive white gaussian noise along with multipath fading. It contains modulated signals with 4~\ac{sps} and a sample length of 128 samples. Used modulations along with the complete parameter list can be found in Table~\ref{table_rml_dataset}. Detailed specifications and generation details of the dataset can be found in \cite{o2016radio}.

\subsection{Modified RadioML dataset}
The standard radioML dataset is extended using the generation code\footnote{https://github.com/radioML/dataset} by varying the \textit{samples per symbol} and \textit{sample length} parameters for evaluating the sample rate dependencies of the \ac{lstm} model. The extended parameters of the used dataset are listed in the Table~\ref{table_modrml_dataset}. The extended dataset contains signals with 4 and 8 samples per symbol. This dataset is generated to evaluate the robustness of the model in varying symbol rate scenarios.

\section{Model Description}
\label{models}
\begin{figure}[htb]
\centering
\includegraphics[width=1\columnwidth]{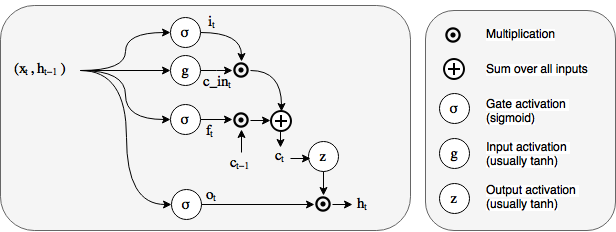}
\caption{LSTM cell used in the hidden layers of the model.} 
\label{fig_lstmcell}
\end{figure}


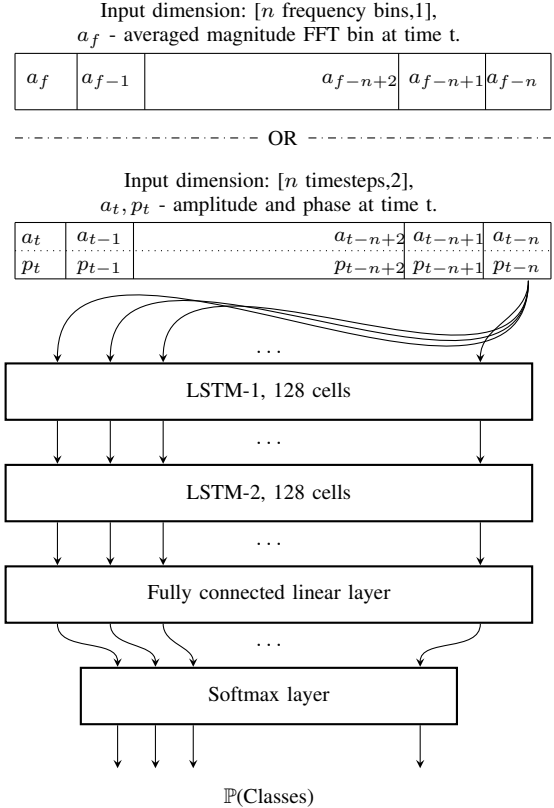
\begin{figure}[t]
\footnotesize
	\centering
\begin{tikzpicture}[x=1.5cm, y=1.5cm, >=stealth]
\draw (-2.75,5.5) -- (2,5.5);
\draw [dotted] (-2.75,5.25) -- (2,5.25);
\draw (-2.75,5) -- (2,5);
\draw (-2.75,5.5) -- (-2.75,5);

\draw (-2.75,7) -- (2,7);
\draw (-2.75,6.5) -- (2,6.5);
\draw (-2.75,7) -- (-2.75,6.5);

\foreach \x/\c in {-2.2/{f}, -1.6/{f-1}, 0.65/{f-n+2}, 1.42/{f-n+1}, 2/{f-n}}
{
    \draw (\x,7) -- (\x,6.5);
    \node at (\x-0.34, 6.75) {${a_{\c}}$};
}
\node (inplogo0) at (-0.5,7.25) [,minimum width=5cm,minimum height=0.55cm, align=center] {Input dimension: [$n$ frequency bins,1],\\ $a_f$ - averaged magnitude FFT bin at time t.};,

\draw [dashdotted] (-2.75,6.25) -- (2,6.25) node[midway,fill=white] {OR} ;
\foreach \x/\c in {-2.3/{t}, -1.7/{t-1}, 0.7/{t-n+2}, 1.4/{t-n+1}, 2/{t-n}}
{
    \draw (\x,5.5) -- (\x,5);
    \node at (\x-0.3, 5.1) {${p_{\c}}$};
    \node at (\x-0.3, 5.35) {${a_{\c}}$};
}  

\node (inplogo) at (-0.5,5.75) [,minimum width=5cm,minimum height=0.55cm, align=center] {Input dimension: [$n$ timesteps,2],\\ $a_t, p_t$ - amplitude and phase at time t.};
\node (inp) at (1.8,5.05) [] {};
\node (lstm1) at (-0.5,4) [draw,thick,minimum width=7cm,minimum height=0.75cm] {\ac{lstm}-1, 128 cells};
\node (lstm2) at (-0.5,3.1) [draw,thick,minimum width=7cm,minimum height=0.75cm] {\ac{lstm}-2, 128 cells};
\node (fc) at (-0.5,2.2) [draw,thick,minimum width=7cm,minimum height=0.75cm] {Fully connected linear layer};
\node (sm) at (-0.5,1.3) [draw,thick,minimum width=5cm,minimum height=0.75cm] {Softmax layer};
\node (dummy) at (-0.5,0.4) [draw=none,thick,minimum width=5cm,minimum height=0.75cm] {$\mathbb{P}$(Classes)};
\foreach \a in {0.1,0.2,0.3,0.9}{
\draw ($(inp.south)$) edge[out=270,in=90,->] ($(lstm1.north west)!\a!(lstm1.north east)$);
\draw[->] ($(lstm1.south west)!\a!(lstm1.south east)$) -- ($(lstm2.north west)!\a!(lstm2.north east)$);
\draw[->] ($(lstm2.south west)!\a!(lstm2.south east)$) -- ($(fc.north west)!\a!(fc.north east)$);
\draw ($(fc.south west)!\a!(fc.south east)$) edge[out=270,in=90,->] ($(sm.north west)!\a!(sm.north east)$);
\draw[->] ($(sm.south west)!\a!(sm.south east)$) -- ($(dummy.north west)!\a!(dummy.north east)$);
}
\node at ($(inplogo)!0.8!(lstm1)$) {$\hdots$};
\node at ($(lstm1)!0.5!(lstm2)$) {$\hdots$};
\node at ($(lstm2)!0.5!(fc)$) {$\hdots$};
\node at ($(fc)!0.5!(sm)$) {$\hdots$};

\end{tikzpicture}
	\caption{Two layer \ac{lstm} model for classification. The model is trained and deployed for modulation classification using either amplitude-phase signal or the averaged magnitude-FFT signal as input.}
	\label{fig_iq_lstm_model}
\end{figure}

The proposed \ac{lstm} model, that works on the time domain amplitude and phase signal, is introduced in the following subsection. In addition, the baseline \ac{cnn} model used for comparisons is also detailed.
\subsection{LSTM primer}
\label{lstm_primer}
\ac{rnn} are heavily used for learning persistent features from time series data. \ac{lstm} \cite{Hochreiter:lstm} is a special type of \ac{rnn} which is efficient in learning long-term dependencies. The block diagram of a basic version of a \ac{lstm} cell is presented in Figure~\ref{fig_lstmcell} along with the corresponding equations (2-7).

\begin{itemize}
\item Gates
\begin{align}
i_t &= \sigma(W_{xi}x_t + W_{hi}h_{t-1} + b_i)\\
f_t &= \sigma(W_{xf}x_t + W_{hf}h_{t-1} + b_f)\\
o_t &= \sigma(W_{xo}x_t + W_{ho}h_{t-1} + b_o)
\end{align}
\item Input transform
\begin{equation}
c\_in_t = tanh(W_{xc}x_t+W_{hc}h_{t-1}+b_{c\_in})
\end{equation}
\item State update
\begin{align}
c_t &= f_t \cdot c_{t-1}+i_t \cdot c\_in_t\\
h_t &= o_t \cdot tanh(c_t)
\end{align}
\end{itemize}

\ac{lstm} cells have an internal state or memory ($c_t$) along with three gates namely input date ($i_t$), forget gate ($f_t$) and output gate ($f_t$). Based on the previous state and the input data the cells can learn the gate weights for the specified problem. This gating mechanism helps \ac{lstm} cells to store information for longer duration thereby enabling persistent feature learning.
\subsection{Model for complex signals}\label{cmplx_lstm_model}
A \ac{lstm} network with different layers is used for complex data classification as shown in Figure \ref{fig_iq_lstm_model}. The amplitude and phase of the time domain modulated signal are fed to all cells of the \ac{lstm} model as a two dimensional vector, at each time step for classification. The amplitude vector is L2 normalized and the phase, which is in radians is normalized between -1 and 1. The first two layers are comprised of 128 \ac{lstm} cells each. The final output from the second \ac{lstm} layer, a vector of dimension 128, after all time steps, is fed to the last dense layer of the model. The final layer is a dense softmax layer which maps the classified features to one of the 11 output classes representing the modulation schemes. The two layer model is selected after detailed analysis varying the cell size and layer depths which are detailed in Section~\ref{sec_hyperparams}. 

The intuition to use a \ac{lstm} model for classification is based on the fact that different modulation schemes exhibit different amplitude and phase characteristics and the model can learn these temporal dependencies effectively. Even though fading and other real world effects may slightly hamper the characteristics of the signal, we expect the model to classify signals efficiently by learning good fading resistant representations. Since the proposed model can work on variable length input time domain samples, we expect the model to learn useful symbol rate independent representations for classification. In addition, the importance of the number of \ac{lstm} cells and layer depth are further investigated by varying these parameters. Model classification accuracies are analyzed with varying layer depth from 1 to 3 and number of cells from 16 to 256. We further analyze these aspects in detail in Section~\ref{results}.

\subsection{Baseline \ac{iq} model}
The two layer CNN 8 tap model presented in \cite{baseline} is used as the baseline model for further comparisons. The baseline model uses 256 and 80 filters in the first two convolutional layers respectively. A publicly available training model is used for generating the baseline performance graph \cite{o2016convolutional}. 

\subsection{Model training and testing}
Each of the datasets mentioned in Tables~\ref{table_rml_dataset} and \ref{table_modrml_dataset} are split into two, one training set and the other testing set. A seed is used to generate random mutually exclusive array indices, which are then used to split the data into two ascertaining the training and testing sets are entirely different. The number of the training and testing vectors are listed in the corresponding tables. A softmax cross entropy with logits\footnote{https://www.tensorflow.org/api\_docs/python/tf/nn/softmax\_cross\_\\entropy\_with\_logits}, that measures the probability error in discrete classification tasks in which the classes are mutually exclusive, is used as the loss function. Stochastic gradient descent with a minibatch size of 400 vectors is used to avoid local optima. We use the Adam optimizer \cite{adam_optimizer}, a first-order gradient based optimizer with a learning rate of 0.001. The complex two layer \ac{lstm} network is trained for 70 epochs which takes around an hour of training time on a x86 PC with Nvidia GeForce GTX 980 Ti graphics card. We use a basic \ac{lstm} cell with starting training forget bias set to one. While initializing the network, it is helpful to keep the scale of the input variance constant, so that it does not explode or diminish by reaching the final layer. To achieve this \ac{lstm} weights are initialized with a default uniform unit scaling initializer which generates weights with a uniform variance. All the models use the same training parameters unless specified explicitly.

\subsection{Implementation details}\label{implementation}
The neural network is implemented using TensorFlow \cite{tensorflow}, a data flow graph based numerical computation library from Google. Python and C++ bindings of Tensorflow makes the usage of the final trained model easily portable to host based SDR frameworks like GNU Radio \cite{gnuradio_web}. The trained model can be easily imported as a block in GNU Radio which can be readily used in practice with any supported hardware front-end.

\section{Results and discussion}
\label{results}
The classification accuracies of the model for the aforementioned datasets along with the learned representations are discussed in the following subsections.

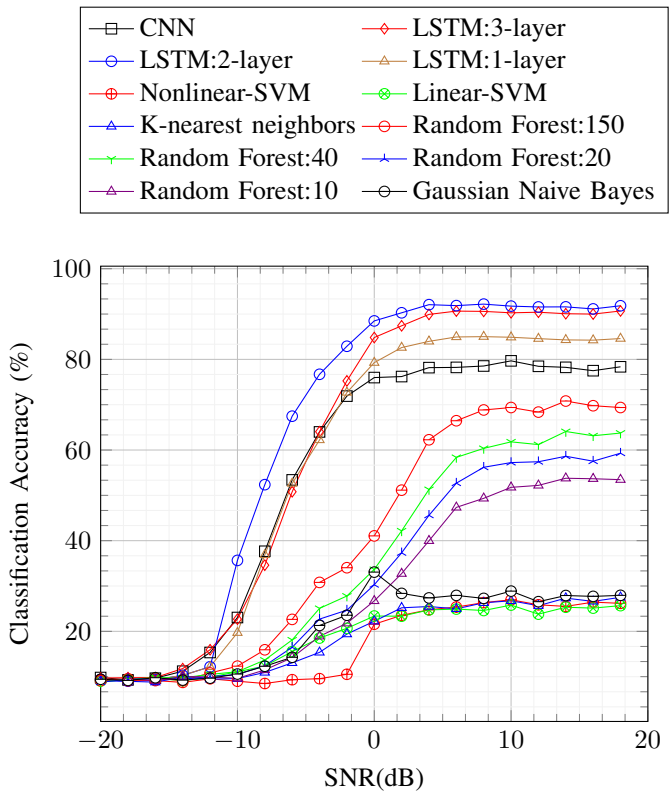
\begin{figure}[htb]
\begin{tikzpicture} \begin{axis}[legend style={at={(0.5,1.57)},anchor=north},
legend columns=2,
 width=\columnwidth,
 grid=both,
 ylabel=Classification Accuracy (\%),
  xlabel=SNR(dB),
 grid style={line width=.1pt, draw=gray!10},
 major grid style={line width=.2pt,draw=gray!50},
 xmin=-20,
 xmax=20,
 minor tick num=5,
 legend cell align={left},
]
\addplot[mark=square, color=black] coordinates { 
(-20, 09.807868252516011) (-18, 09.282470481380563) (-16, 09.659913169319827) (-14, 11.205341032118368) (-12, 15.332725615314494) (-10, 23.049582370712635) (-8, 37.6630534631637) (-6, 53.40909090909091) (-4, 63.99260628465804) (-2, 71.89767779390421) (0, 75.98540145985402) (2, 76.20865139949109) (4, 78.1754772393539) (6, 78.23104693140794) (8, 78.53922452660054) (10, 79.67213114754098) (12, 78.47184501176897) (14, 78.2432183059605) (16, 77.50226244343892) (18, 78.36183618361836) };
\addlegendentry{CNN}
\addplot[mark=diamond, color=red] coordinates { 
( -20 , 9.69807868253 )  ( -18 , 9.77293369664 )  ( -16 , 9.80463096961 )  ( -14 , 11.7827499098 )  ( -12 , 15.934366454 )  ( -10 , 22.69415319 )  ( -8 , 34.5948925225 )  ( -6 , 50.7881231672 )  ( -4 , 64.0480591497 )  ( -2 , 75.2539912917 )  ( 0 , 84.8175182482 )  ( 2 , 87.4045801527 )  ( 4 , 89.922907489 )  ( 6 , 90.6498194946 )  ( 8 , 90.5861136159 )  ( 10 , 90.2732240437 )  ( 12 , 90.4037660692 )  ( 14 , 90.071968998 )  ( 16 , 89.9909502262 )  ( 18 , 90.6750675068 )
};
\addlegendentry{LSTM:3-layer}
\addplot[mark=o, color=blue] coordinates {
(-20, 09.387008234217749) (-18, 08.991825613079019) (-16, 09.334298118668596) (-14, 10.285095633345363) (-12, 12.178669097538743) (-10, 35.64954682779456) (-8,  52.36083042439831) (-6,  67.46700879765396) (-4,  76.7097966728281) (-2,  82.89187227866474) (0, 88.48540145985402) (2, 90.27626317702654) (4, 92.0704845814978) (6, 91.85920577617328) (8, 92.19116321009919) (10, 91.74863387978142) (12, 91.56255658156799) (14, 91.58516331426463) (16, 91.13122171945701) (18, 91.82718271827183)
};
\addlegendentry{LSTM:2-layer}
\addplot[mark=triangle, color=brown] coordinates {
(-20, 9.3870082342177488) (-18, 8.9918256130790186) (-16, 9.3342981186685964) (-14,  10.285095633345363) (-12, 12.178669097538743) (-10,19.699166364624865) (-8, 36.863319897398317) (-6, 52.61261261261261) (-4, 62.22222222222222) (-2, 72.69068959209805) (0,  79.27895120174799) (2,  82.60468548238332) (4,  83.97647923557515) (6,  84.94662565587118) (8,  85.01545735588288) (10, 84.88180318856514) (12, 84.57747764190545) (14, 84.29677651719791) (16, 84.23833819241983) (18, 84.59328425210989)
};
\addlegendentry{LSTM:1-layer}
\addplot[mark=oplus, color=red] coordinates {
(  -20 , 9.29551692589 )(  -18 , 9.10081743869 )(  -16 , 9.13531114327 )(  -14 , 8.73330927463 )(  -12 , 9.53509571559 )(  -10 , 8.99235827261 )(  -8 , 8.48796619511 )(  -6 , 9.32917888563 )(  -4 , 9.55637707948 )(  -2 , 10.5224963716 )(  0 , 21.5693430657 )(  2 , 23.5368956743 )(  4 , 24.7613803231 )(  6 , 25.3610108303 )(  8 , 26.3300270514 )(  10 , 26.9581056466 )(  12 , 25.855513308 )(  14 , 25.5028603063 )(  16 , 26.443438914 )(  18 , 26.1746174617 )
};
\addlegendentry{Nonlinear-SVM}
\addplot[mark=otimes, color=green] coordinates {
(  -20 , 9.02104300091 )(  -18 , 9.42779291553 )(  -16 , 9.62373371925 )(  -14 , 9.1122338506 )(  -12 , 9.89972652689 )(  -10 , 11.0716189799 )(  -8 , 12.3461326474 )(  -6 , 15.5241935484 )(  -4 , 18.5212569316 )(  -2 , 20.6640058055 )(  0 , 23.3759124088 )(  2 , 23.3369683751 )(  4 , 24.7246696035 )(  6 , 24.9097472924 )(  8 , 24.5987376014 )(  10 , 25.7923497268 )(  12 , 23.7914177078 )(  14 , 25.3183244141 )(  16 , 25.1221719457 )(  18 , 25.6705670567 )
};
\addlegendentry{Linear-SVM}
\addplot[mark=triangle, color=blue] coordinates {
(  -20 , 9.09423604758 )(  -18 , 8.97366030881 )(  -16 , 8.79160636758 )(  -14 , 9.92421508481 )(  -12 , 9.44393801276 )(  -10 , 9.57881642083 )(  -8 , 10.9314716149 )(  -6 , 13.0498533724 )(  -4 , 15.3604436229 )(  -2 , 19.3940493469 )(  0 , 22.2262773723 )(  2 , 25.1908396947 )(  4 , 25.4772393539 )(  6 , 24.963898917 )(  8 , 26.2939585212 )(  10 , 26.7395264117 )(  12 , 25.6563461887 )(  14 , 27.3297656394 )(  16 , 26.6063348416 )(  18 , 27.5067506751 )
};
\addlegendentry{K-nearest neighbors}
\addplot[mark=halfcircle, color=red] coordinates {
(  -20 , 9.11253430924 )(  -18 , 9.55495004541 )(  -16 , 9.65991316932 )(  -14 , 9.14832190545 )(  -12 , 10.8113035552 )(  -10 , 12.3156211125 )(  -8 , 15.9287157817 )(  -6 , 22.6356304985 )(  -4 , 30.7948243993 )(  -2 , 34.0348330914 )(  0 , 41.0583941606 )(  2 , 51.1450381679 )(  4 , 62.2613803231 )(  6 , 66.4620938628 )(  8 , 68.8367899008 )(  10 , 69.3806921676 )(  12 , 68.3686402318 )(  14 , 70.806421849 )(  16 , 69.7737556561 )(  18 , 69.3789378938 )
};
\addlegendentry{Random Forest:150}
\addplot[mark=Mercedes star flipped, color=green] coordinates {
(  -20 , 8.78316559927 )(  -18 , 9.4459582198 )(  -16 , 9.76845151954 )(  -14 , 9.74377481054 )(  -12 , 10.5013673655 )(  -10 , 11.089390439 )(  -8 , 13.6689325739 )(  -6 , 18.1085043988 )(  -4 , 25.0646950092 )(  -2 , 27.793904209 )(  0 , 33.6678832117 )(  2 , 42.1664849146 )(  4 , 51.3032305433 )(  6 , 58.3754512635 )(  8 , 60.3606853021 )(  10 , 61.8032786885 )(  12 , 61.1986239363 )(  14 , 64.0893153718 )(  16 , 63.149321267 )(  18 , 63.7443744374 )
};
\addlegendentry{Random Forest:40}
\addplot[mark=Mercedes star, color=blue] coordinates {
(  -20 , 8.7648673376 )(  -18 , 9.42779291553 )(  -16 , 9.53328509407 )(  -14 , 9.45507037171 )(  -12 , 10.0091157703 )(  -10 , 10.4673893727 )(  -8 , 12.4196215322 )(  -6 , 16.4772727273 )(  -4 , 22.6987060998 )(  -2 , 24.6734397678 )(  0 , 30.2737226277 )(  2 , 37.3137041076 )(  4 , 45.5947136564 )(  6 , 52.6895306859 )(  8 , 56.1947700631 )(  10 , 57.2313296903 )(  12 , 57.4144486692 )(  14 , 58.6270529618 )(  16 , 57.5384615385 )(  18 , 59.297929793 )
};
\addlegendentry{Random Forest:20}
\addplot[mark=triangle, color=violet] coordinates {
(  -18 , 9.57311534968 )(  -16 , 9.08104196816 )(  -14 , 10.0505232768 )(  -12 , 9.91795806746 )(  -10 , 9.63213079794 )(  -8 , 11.482638251 )(  -6 , 14.2045454545 )(  -4 , 18.9094269871 )(  -2 , 21.7888243832 )(  0 , 26.7153284672 )(  2 , 32.7335514358 )(  4 , 39.9779735683 )(  6 , 47.3465703971 )(  8 , 49.3417493237 )(  10 , 51.766848816 )(  12 , 52.1998913634 )(  14 , 53.7737589961 )(  16 , 53.665158371 )(  18 , 53.4653465347 )
};
\addlegendentry{Random Forest:10}
\addplot[mark=halfcircle, color=black] coordinates {
(  -20 , 9.42360475755 )(  -18 , 9.17347865577 )(  -16 , 9.76845151954 )(  -14 , 9.23854204258 )(  -12 , 9.66271649954 )(  -10 , 10.5740181269 )(  -8 , 12.2910159838 )(  -6 , 14.2778592375 )(  -4 , 21.2939001848 )(  -2 , 23.5486211901 )(  0 , 33.0656934307 )(  2 , 28.3715012723 )(  4 , 27.3678414097 )(  6 , 27.9241877256 )(  8 , 27.2678088368 )(  10 , 28.8706739526 )(  12 , 26.5435451747 )(  14 , 27.9018269053 )(  16 , 27.7104072398 )(  18 , 27.9387938794 )
};
\addlegendentry{Gaussian Naive Bayes}
\addplot[mark=triangle, color=brown] coordinates {
};
\addlegendentry{MNB}
\addplot[mark=triangle, color=brown] coordinates {

};
\addlegendentry{SVC}
\end{axis}
\end{tikzpicture}
\caption{Classification accuracy comparison of hyper-parameter optimized 2-layer amplitude-phase \ac{lstm} model with others on RadioML dataset.}
\label{fig_lstm_rml_acc}
\end{figure}

\begin{figure}[htb]
\centering
-\includegraphics[width=1.1\columnwidth]{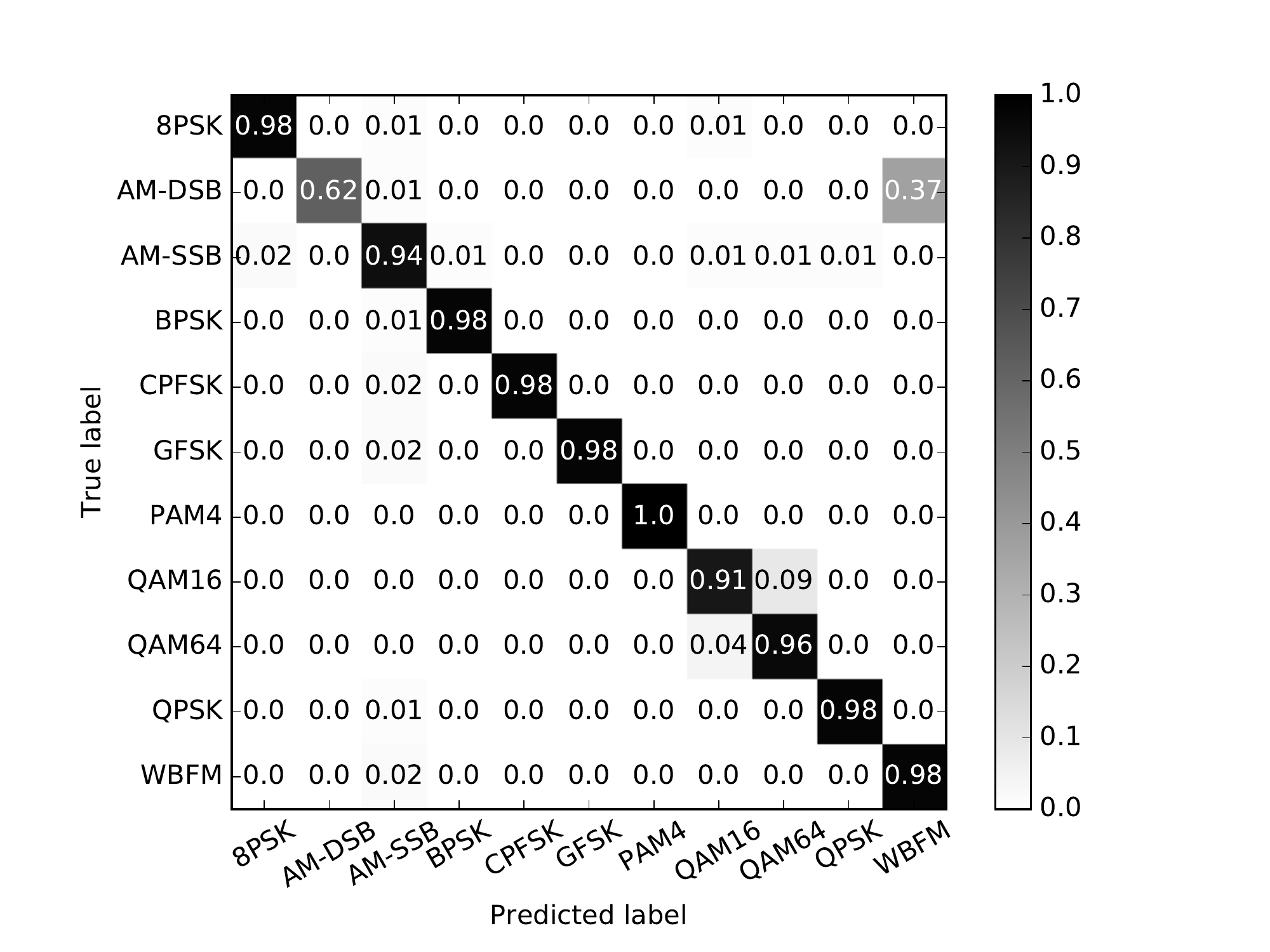}
\caption{Confusion matrix for 2-layer amplitude-phase LSTM model on RadioML dataset at 18dB SNR} 
\label{fig_confmat_18}
\end{figure}

\begin{figure}[htb]
\centering
\includegraphics[width=1.1\columnwidth]{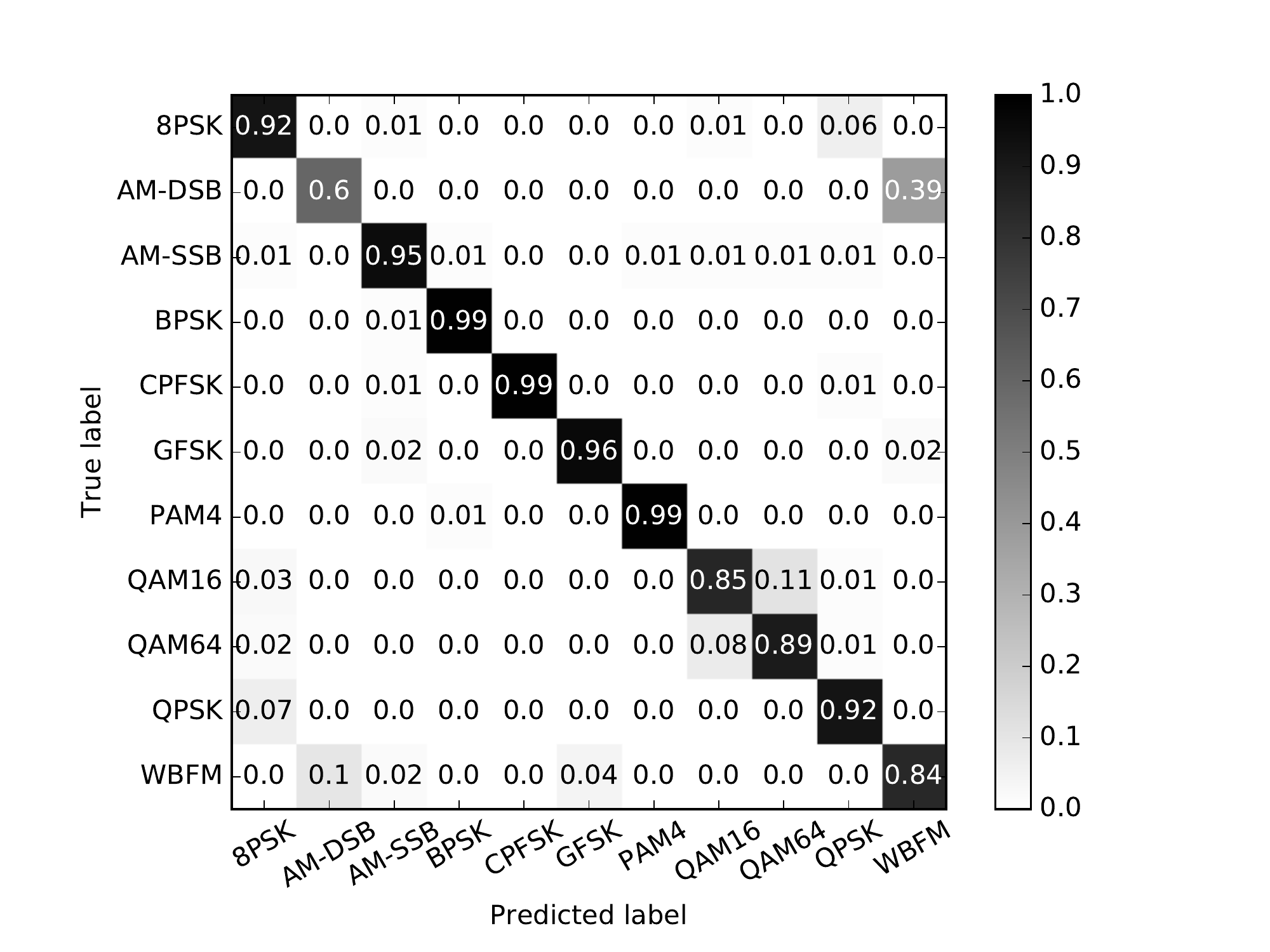}
\caption{Confusion matrix for 2-layer amplitude-phase LSTM model on RadioML dataset at 0dB SNR} 
\label{fig_confmat_0}
\end{figure}

\begin{figure}[htb]
\centering
\squeezeup
\includegraphics[width=1.1\columnwidth]{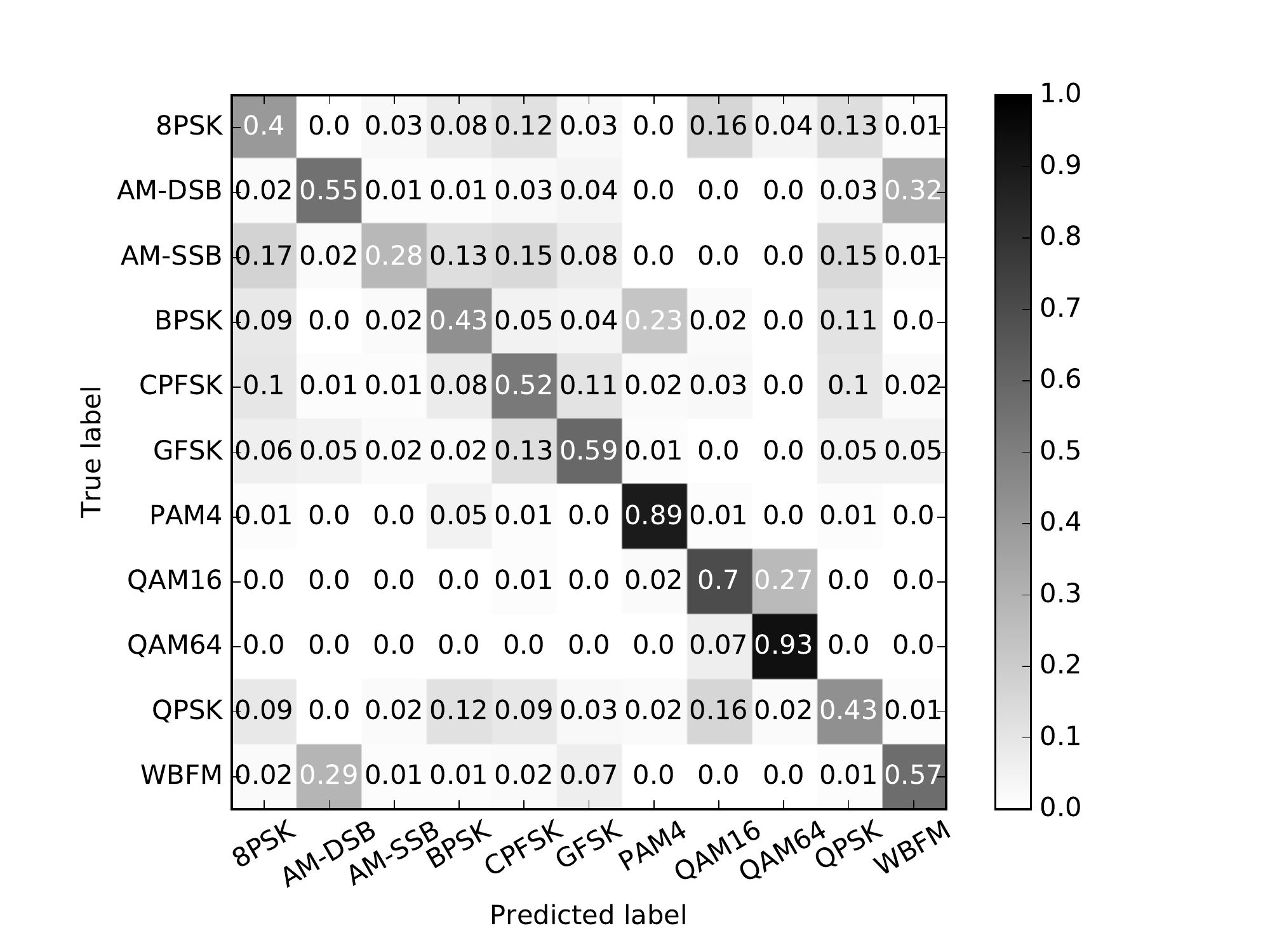}
\caption{Confusion matrix for 2-layer amplitude-phase LSTM model on RadioML dataset at -8dB SNR} 
\label{fig_confmat_-8}
\end{figure}

\begin{figure}[htb]
\begin{tikzpicture} \begin{axis}[legend pos=south east,
 width=\columnwidth,
 grid=both,
 ylabel=Classification Accuracy (\%),
  xlabel=SNR(dB),
 grid style={line width=.1pt, draw=gray!10},
 major grid style={line width=.2pt,draw=gray!50},
 minor tick num=5,
 xmin=-20,
 xmax=20,
 legend cell align={left},
]
\addplot[mark=square, color=black] coordinates {
(-20, 09.405306495882891) (-18, 09.137148047229791) (-16, 09.225759768451519) (-14, 09.978347167087694) (-12, 15.478577939835916) (-10, 28.45210591789586) (-8,  44.497519750137793) (-6,  58.77932551319648) (-4,  72.60628465804067) (-2,  83.30914368650217) (0,   87.40875912408759) (2,   89.96728462377317) (4,   91.70337738619677) (6,   92.49097472924188) (8,   93.86834986474302) (10,  93.04189435336976) (12,  93.26453014665942) (14,  92.96918250599742) (16,  92.66968325791856) (18,  93.01530153015302)
};
\addlegendentry{8sps, n=512 samples}
\addplot[mark=o, color=blue] coordinates {
(-20, 09.29551692589204) (-18, 09.409627611262489) (-16, 09.370477568740955) (-14, 10.267051605918441) (-12, 14.62169553327256) (-10, 23.991469699662343) (-8,  38.17747565680691) (-6,  52.74926686217009) (-4,  67.8003696857671) (-2,  80.76923076923077) (0,   85.62043795620438) (2,   87.94983642311887) (4,   89.97797356828194) (6,   90.41516245487364) (8,   91.09107303877367) (10,  90.65573770491804) (12,  91.01937352887923) (14,  90.55176231777081) (16,  91.23981900452489) (18,  90.72907290729073)
};
\addlegendentry{8sps, n=256 samples}
\addplot[mark=triangle, color=red] coordinates {
(-20, 09.09423604757548) (-18, 09.336966394187103) (-16, 09.659913169319827) (-14, 10.230963551064598) (-12, 12.725615314494074) (-10, 18.393460103074463) (-8,  28.32996509277972) (-6,  43.603372434017595) (-4,  57.7818853974122) (-2,  71.31712626995645) (0,   79.76277372262773) (2,   83.33333333333334) (4,   85.11380323054332) (6,   86.15523465703971) (8,   87.14156898106402) (10,  86.61202185792349) (12,  86.8730762266884) (14,  85.90145783354862) (16,  86.53393665158371) (18,  86.13861386138614)
};
\addlegendentry{8sps, n=128 samples}
\addplot[dashed, mark=diamond, color=brown] coordinates {
(-20, 09.149130832570906) (-18, 09.30063578564941) (-16, 09.479015918958032) (-14, 09.906171057380007) (-12, 12.123974475843209) (-10, 14.448196196907767) (-8,  20.65037663053463) (-6,  30.97507331378299) (-4,  43.21626617375231) (-2,  51.94121915820029) (0,   61.77007299270073) (2,   65.13994910941476) (4,   67.7863436123348) (6,   68.35740072202167) (8,   68.85482416591524) (10,  68.70673952641165) (12,  68.45917074053957) (14,  67.66931168112198) (16,  69.1764705882353) (18,  69.23492349234923)
};
\addlegendentry{8sps, n=64 samples}
\end{axis}
\end{tikzpicture}
\caption{Classification accuracy of amplitude-phase \ac{lstm} model on modified RadioML dataset for 8 samples/symbol. The model is trained only on input sample lengths (n) from 128 to 512. The model gives close to 70\% accuracy on 64 input samples (dashed line) which it is not trained for. }
\label{fig_lstm_modrml_acc_8sps}
\end{figure}

\begin{figure}[htb]
\begin{tikzpicture} \begin{axis}[legend pos=south east,
 width=\columnwidth,
 grid=both,
 ylabel=Classification Accuracy (\%),
  xlabel=SNR(dB),
 grid style={line width=.1pt, draw=gray!10},
 major grid style={line width=.2pt,draw=gray!50},
 minor tick num=5,
 xmin=-20,
 xmax=20,
 legend cell align={left},
]
\addplot[mark=square, color=black] coordinates {
(-20, 09.075937785910339) (-18, 09.30063578564941) (-16, 09.280028943560058) (-14, 09.870083002526164) (-12, 15.460346399270739) (-10, 26.852674604585036) (-8,  38.875620062465555) (-6,  51.55791788856305) (-4,  62.64325323475046) (-2,  77.99346879535559) (0,   86.45985401459854) (2,   88.38604143947656) (4,   90.45521292217328) (6,   92.03971119133574) (8,   92.89449954914337) (10,  92.89617486338798) (12,  93.64475828354155) (14,  92.54474995386602) (16,  93.90045248868778) (18,  93.06930693069307)
};
\addlegendentry{4sps, n=512 samples}
\addplot[mark=o, color=blue] coordinates {
(-20, 09.277218664226898) (-18, 09.282470481380563) (-16, 09.352387843704775) (-14, 10.230963551064598) (-12, 14.275296262534184) (-10, 23.867069486404835) (-8,  34.41117031049054) (-6,  48.277126099706746) (-4,  60.03696857670979) (-2,  74.92743105950653) (0,   84.1970802919708) (2,   86.93202471828426) (4,   89.17033773861968) (6,   89.98194945848376) (8,   91.36158701532913) (10,  91.94899817850638) (12,  92.17816404128191) (14,  92.04650304484222) (16,  92.50678733031674) (18,  91.8091809180918)
};
\addlegendentry{4sps, n=256 samples}
\addplot[mark=triangle, color=red] coordinates {
(-20, 09.222323879231473) (-18, 09.30063578564941) (-16, 09.460926193921852) (-14, 09.906171057380007) (-12, 12.61622607110301) (-10, 18.340145725964102) (-8,  27.246004041888666) (-6,  38.41642228739003) (-4,  52.80961182994455) (-2,  66.16473149492017) (0,   78.37591240875912) (2,   83.6423118865867) (4,   86.04992657856094) (6,   87.76173285198556) (8,   88.4400360685302) (10,  88.45173041894353) (12,  89.10012674271229) (14,  88.78021775235283) (16,  88.83257918552037) (18,  88.17281728172818)
};
\addlegendentry{4sps, n=128 samples}
\addplot[dashed, mark=diamond, color=brown] coordinates {
(-20, 08.947849954254346) (-18, 08.810172570390554) (-16, 08.990593342981186) (-14, 09.455070371706965) (-12, 12.069279854147676) (-10, 14.643682246312423) (-8,  20.687121072937717) (-6,  29.30718475073314) (-4,  40.499075785582256) (-2,  51.37880986937591) (0,   63.17518248175182) (2,   69.08396946564885) (4,   71.51248164464024) (6,   74.02527075812274) (8,   74.35527502254283) (10,  73.64298724954462) (12,  74.72388194821655) (14,  74.46023251522421) (16,  74.28054298642534) (18,  74.97749774977498)
};
\addlegendentry{4sps, n=64 samples}
\end{axis}
\end{tikzpicture}
\caption{Classification accuracy  of amplitude-phase \ac{lstm} model on modified RadioML dataset for 4 samples/symbol. The model is trained only on input sample lengths (n) from 128 to 512. The model gives close to 75\% accuracy on 64 input samples (dashed line) which it is not trained for.}
\label{fig_lstm_modrml_acc_4sps}
\end{figure}

\begin{figure}[htb]
\begin{tikzpicture} \begin{axis}[legend pos=south east,
 width=\columnwidth,
 grid=both,
 ylabel=Classification Accuracy (\%),
  xlabel=SNR(dB),
 grid style={line width=.1pt, draw=gray!10},
 major grid style={line width=.2pt,draw=gray!50},
 minor tick num=5,
 xmin=-20,
 xmax=20,
 ymax=100,
 legend cell align={left},
]

\addplot[mark=square, color=black] coordinates {
(  -20 , 9.29551692589 )(  -18 , 9.19164396004 )(  -16 , 9.20767004342 )(  -14 , 8.76939732948 )(  -12 , 9.88149498633 )(  -10 , 10.14750311 )(  -8 , 12.0154326658 )(  -6 , 18.7316715543 )(  -4 , 25.6561922366 )(  -2 , 44.3759071118 )(  0 , 57.6459854015 )(  2 , 59.9418393312 )(  4 , 66.5198237885 )(  6 , 69.3140794224 )(  8 , 70.3877366997 )(  10 , 71.693989071 )(  12 , 70.2516748144 )(  14 , 70.9171433844 )(  16 , 70.1719457014 )(  18 , 71.2871287129 )
};
\addlegendentry{4sps, n=512 samples}
\addplot[mark=o, color=blue] coordinates {
(  -20 , 9.22232387923 )(  -18 , 8.99182561308 )(  -16 , 9.58755426918 )(  -14 , 9.76181883796 )(  -12 , 11.7046490428 )(  -10 , 13.3819086547 )(  -8 , 17.1963990446 )(  -6 , 22.6173020528 )(  -4 , 35.6561922366 )(  -2 , 52.8301886792 )(  0 , 59.1240875912 )(  2 , 66.2849872774 )(  4 , 69.7320117474 )(  6 , 70.6859205776 )(  8 , 70.441839495 )(  10 , 70.1275045537 )(  12 , 69.5636429477 )(  14 , 70.6957003137 )(  16 , 70.8416289593 )(  18 , 70.7110711071 )
};
\addlegendentry{4sps, n=256 samples}
\addplot[mark=triangle, color=red] coordinates {
(  -20 , 9.2406221409 )(  -18 , 9.48228882834 )(  -16 , 9.47901591896 )(  -14 , 9.99639119451 )(  -12 , 11.5587967183 )(  -10 , 14.2704816065 )(  -8 , 18.9968767224 )(  -6 , 27.5843108504 )(  -4 , 43.1423290203 )(  -2 , 56.8396226415 )(  0 , 64.7262773723 )(  2 , 69.7746274082 )(  4 , 73.4214390602 )(  6 , 74.4945848375 )(  8 , 74.6798917944 )(  10 , 75.5919854281 )(  12 , 74.8506246605 )(  14 , 74.9031186566 )(  16 , 75.2941176471 )(  18 , 75.8595859586 )

};
\addlegendentry{4sps, n=128 samples}
\addplot[mark=diamond, color=brown] coordinates {
(  -20 , 9.5516925892 )(  -18 , 9.40962761126 )(  -16 , 9.22575976845 )(  -14 , 10.5016239625 )(  -12 , 12.5068368277 )(  -10 , 14.7680824596 )(  -8 , 19.9706044461 )(  -6 , 26.9978005865 )(  -4 , 37.4676524954 )(  -2 , 48.1494920174 )(  0 , 58.3941605839 )(  2 , 65.0672482734 )(  4 , 69.1997063142 )(  6 , 71.2093862816 )(  8 , 73.3453561767 )(  10 , 73.7704918033 )(  12 , 74.072062285 )(  14 , 73.7959033032 )(  16 , 74.8054298643 )(  18 , 74.3474347435 )
};
\addlegendentry{4sps, n=64 samples}

\end{axis}
\end{tikzpicture}
\caption{Classification accuracy  of amplitude-phase \ac{lstm} model for non-trained data lengths on modified RadioML dataset with 4 samples/symbol.}
\label{fig_lstm_modrml_generalization}
\end{figure}
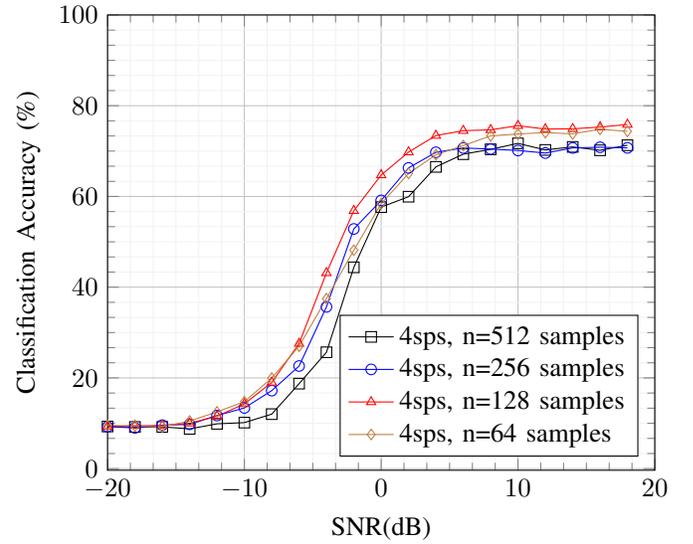

\begin{figure*}[!t]
\centering
\includegraphics[width=1\columnwidth]{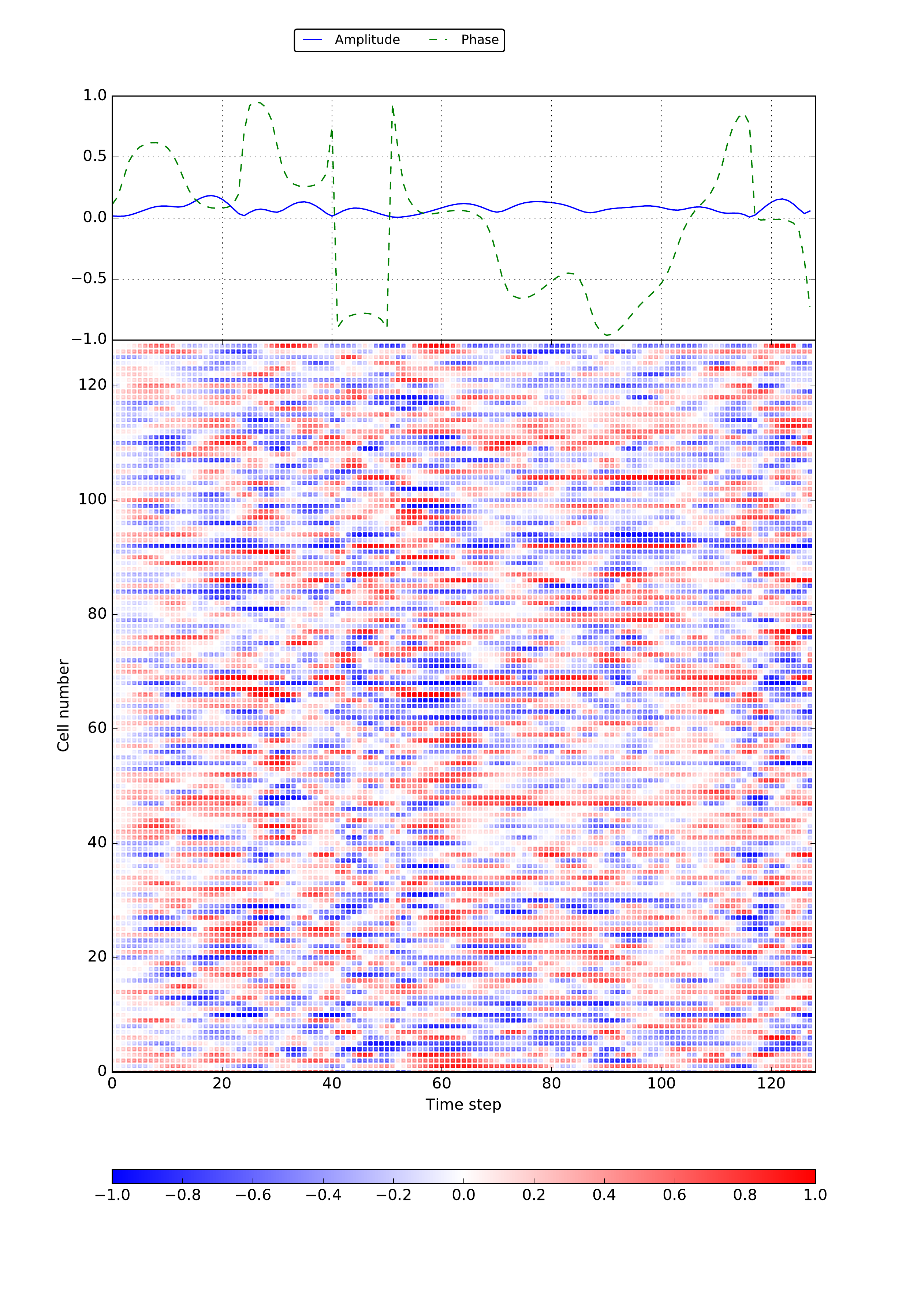}
\includegraphics[width=1\columnwidth]{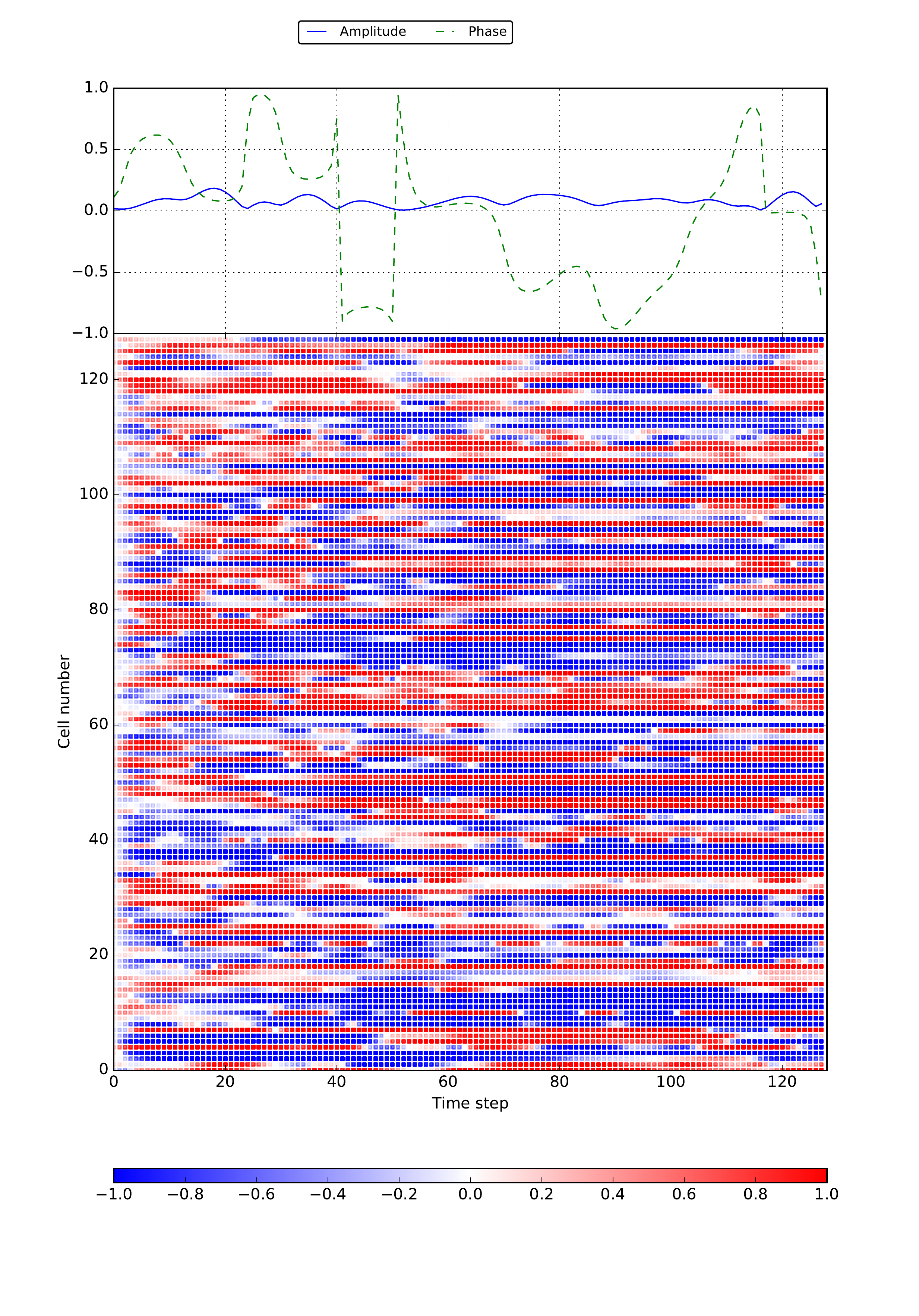}
\caption{Layer-1 (left) and layer-2 (right) LSTM temporal activations for a QAM64 input vector. On the top the amplitude and phase of the input signal (y-axis) is plotted at each time step (x-axis). Below the amplitude and phase of the signal, temporal activations for all cells in the \ac{lstm} model for each time step are shown. Blue denotes \textit{tanh(c)} activations of value -1 and red denotes a value of +1.} 
\label{fig_activations}
\end{figure*}

\begin{figure*}[!t]
\centering
\includegraphics[width=2\columnwidth]{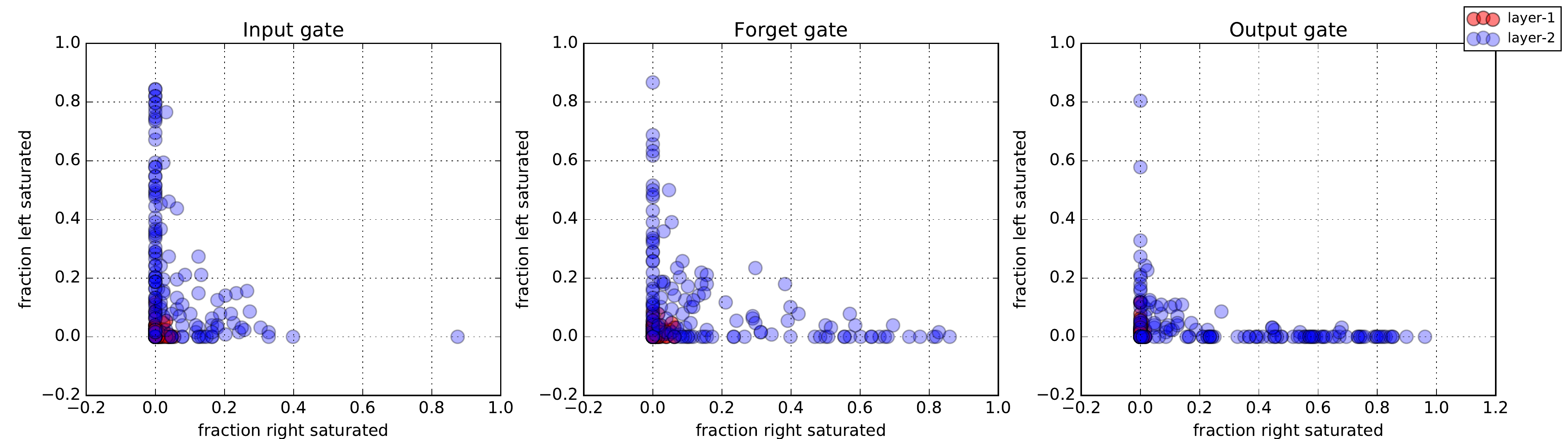}
\caption{Gate saturation plots for LSTM for the same a QAM64 input vector. A circle represents a gate in a particular LSTM cell.} 
\label{fig_gatesat}
\end{figure*}

\subsection{Classification accuracy on RadioML dataset}
The two layer amplitude-phase \ac{lstm} model, shown in Figure~\ref{fig_iq_lstm_model}, is trained on SNR ranges from -10dB to 20dB. Training vectors with SNR ranges below -10dB were not used as the model was converging slowly when those vectors were used. Alternate models with varying \ac{lstm} layer depths are also trained to understand the performance improvements provided by the different layer depths. The classification accuracy of all the four models are presented in Figure~\ref{fig_lstm_rml_acc}. The two layer \ac{lstm} model gave an average accuracy of 90\% in \ac{snr} ranges from 0dB to 20dB. It can be noticed that the single layer \ac{lstm} also reaches a high accuracy at high \ac{snr}s, 6\% less than the two layer model. It was also noticed that the classification accuracy saturates for layer depths of two. Hence, layer depth of two is selected for the final model and its parameters are fine tuned (dropout = 0.8 and learning rate = 0.001) to achieve the best test performance as shown in Figure~\ref{fig_lstm_rml_acc}. Rigorous fine tuning was not performed on layer depths other than two accounting for a slightly lower accuracy levels, for instance the accuracy level for layer depth 3 is slightly lower than layer depth two.  The performance of the baseline \ac{cnn} model was shown to be much better on the low \ac{snr} regions in \cite{baseline}. We were not able to reproduce the reported results on the low \ac{snr} regions after various attempts, which may be because of the difference in hyper-parameter tuning. Though, the high \ac{snr} results of the baseline model matches with that of the reported ones in the paper. Detailed discussions on the effect of layer depth and number of \ac{lstm} cells are presented in Section~\ref{sec_hyperparams}.

Classification performance of other standard machine learning models such as \ac{svm}, random forest, k-nearest neighbors and Gaussian Naive Bayes are also summarized in Figure~\ref{fig_lstm_rml_acc}. All models are fed with the same amplitude-phase training and test data for this comparison. Random forest with 150 decision trees is able to provide close to 70\% of accuracy at very high \ac{snr} conditions while others could reach only around 26\%. It could be clearly noticed that the deep learning models perform superior to the other standard techniques when fed with the raw sensed data. The deep learning models can classify signals very efficiently with a very low number of symbols, usually with hundreds of samples (tens of modulated symbols) when compared to the classical cyclostationary based expert feature models which requires samples in thousands range (hundreds of modulated symbols) for averaging. Similarly extracting expert cyclostationary features using tens of symbols is very suboptimal, which substantiate the use of deep learning models.

To understand the results better confusion matrices for the two layer \ac{lstm} model for various \ac{snr}s are also included. It can be seen in Figure~\ref{fig_confmat_18} that at a high \ac{snr} of 18dB the diagonal is much more sharp even though there are difficulties in separating AM-DSB and WBFM signals. This is mainly due to the silence periods of audio as the modulated signals are generated from real audio streams.  Similarly in Figure~\ref{fig_confmat_0}, at 0dB \ac{snr} it is noticed that there is some level of confusion regarding QAM16 and QAM64 as the former is a subset of the the latter. The confusion increases further at low \ac{snr}s as shown in Figure~\ref{fig_confmat_-8}. From these basic analysis it is clear that deep complex structures as mentioned in \cite{baseline} are not required to achieved good \ac{soa} classification accuracy at high \ac{snr}s. However, use of convolutional layers might turn useful at low \ac{snr}s as reported in \cite{baseline}. In our experiments we also noticed that simply providing \ac{iq} samples to the \ac{lstm} model yielded poor results while normalized amplitude and phase interpretation provided good results. The models even failed to reduce the training loss when fed with time domain \ac{iq} samples, giving a constant accuracy of 9\% on the radioML dataset, as the \ac{lstm}s were not able to extract any meaningful representations. Similarly feeding amplitude-phase information to the \ac{cnn} model did not provide any accuracy improvements over the \ac{iq}-\ac{cnn} model. The classification accuracy improvement is achieved from the combined benefits of using amplitude-phase information along with 2-layer \ac{lstm} model.

\subsection{Classification accuracy on modified RadioML dataset}
The same two layer \ac{lstm} model is trained on SNRs ranging from -20dB to 20dB and input sample lengths from 128 to 512 samples. The accuracy of the model is tested on the full range of SNRs and also on input sample length that is smaller than the training set (e.g, 64). It is evident from the results in Figures~\ref{fig_lstm_modrml_acc_8sps} and \ref{fig_lstm_modrml_acc_4sps} that the classification accuracy improves as the model sees more modulated symbols. Even though the model is trained on varying data lengths from 128 to 512 samples, it gives an average accuracy of 75\% with 64 samples and 4 samples per symbol scenario for which it was not trained, which confirms the model's generalization capabilities. To further analyze the generalization capabilities of the model on unseen sample lengths, four balanced folds of data each containing sequences with sample lengths of 64, 128, 256 and 512 are created. The model is then trained only on three folds, and the left-out fold is used to test generalization to the unseen length. This process is repeated for all four sample lengths and the results are presented in Figure~\ref{fig_lstm_modrml_generalization}. The model consistently gives an average accuracy above 70\% for high SNR conditions. 

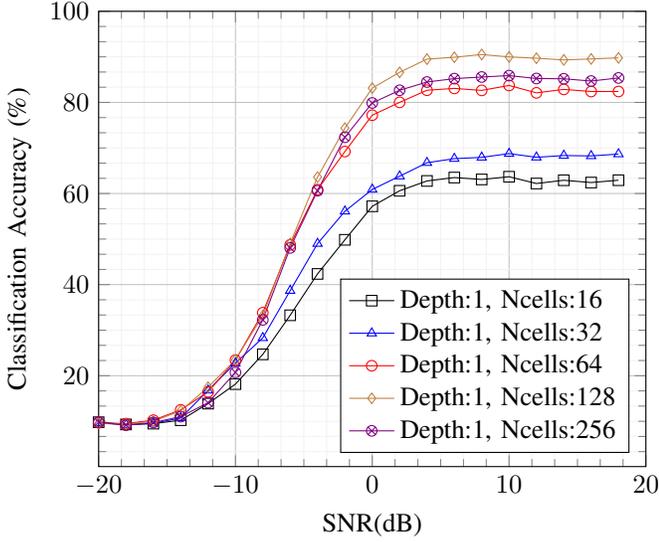
\begin{figure}[htb]
\begin{tikzpicture} \begin{axis}[legend pos=south east,
 width=\columnwidth,
 grid=both,
 ylabel=Classification Accuracy (\%),
  xlabel=SNR(dB),
 grid style={line width=.1pt, draw=gray!10},
 major grid style={line width=.2pt,draw=gray!50},
 minor tick num=5,
 xmin=-20,
 xmax=20,
 ymax=100,
 legend cell align={left},
]

\addplot[mark=square, color=black] coordinates { 
( -20 , 9.78956999085 )  ( -18 , 9.30063578565 )  ( -16 , 9.47901591896 )  ( -14 , 10.2309635511 )  ( -12 , 13.8742023701 )  ( -10 , 18.1624311356 )  ( -8 , 24.6922652949 )  ( -6 , 33.2661290323 )  ( -4 , 42.3475046211 )  ( -2 , 49.8367198839 )  ( 0 , 57.1897810219 )  ( 2 , 60.5961468557 )  ( 4 , 62.7569750367 )  ( 6 , 63.5018050542 )  ( 8 , 63.0838593327 )  ( 10 , 63.679417122 )  ( 12 , 62.1944595329 )  ( 14 , 62.9082856616 )  ( 16 , 62.407239819 )  ( 18 , 62.9162916292 )
};
\addlegendentry{Depth:1, Ncells:16}
\addplot[mark=triangle, color=blue] coordinates { 
( -20 , 9.66148215919 )  ( -18 , 9.20980926431 )  ( -16 , 9.53328509407 )  ( -14 , 10.7181522916 )  ( -12 , 16.7912488605 )  ( -10 , 22.7652390261 )  ( -8 , 28.3115928716 )  ( -6 , 38.6730205279 )  ( -4 , 49.0018484288 )  ( -2 , 56.0957910015 )  ( 0 , 60.8941605839 )  ( 2 , 63.7949836423 )  ( 4 , 66.7584434655 )  ( 6 , 67.6534296029 )  ( 8 , 67.9350766456 )  ( 10 , 68.7613843352 )  ( 12 , 67.9521998914 )  ( 14 , 68.3336408932 )  ( 16 , 68.2533936652 )  ( 18 , 68.6588658866 ) 
};
\addlegendentry{Depth:1, Ncells:32}
\addplot[mark=o, color=red] coordinates { 
( -20 , 9.66148215919 )  ( -18 , 9.48228882834 )  ( -16 , 10.2387843705 )  ( -14 , 12.5045110069 )  ( -12 , 16.4448495898 )  ( -10 , 23.4227830105 )  ( -8 , 33.8416314532 )  ( -6 , 48.7353372434 )  ( -4 , 60.8502772643 )  ( -2 , 69.2126269956 )  ( 0 , 77.1897810219 )  ( 2 , 80.0072700836 )  ( 4 , 82.6725403818 )  ( 6 , 83.0685920578 )  ( 8 , 82.6330027051 )  ( 10 , 83.7158469945 )  ( 12 , 82.093065363 )  ( 14 , 82.8566156117 )  ( 16 , 82.3891402715 )  ( 18 , 82.3942394239 ) 
};
\addlegendentry{Depth:1, Ncells:64}
\addplot[mark=diamond, color=brown] coordinates { 
( -20 , 9.51509606587 )  ( -18 , 9.37329700272 )  ( -16 , 10.003617945 )  ( -14 , 12.3060267052 )  ( -12 , 17.484047402 )  ( -10 , 23.4938688466 )  ( -8 , 33.1986037112 )  ( -6 , 49.0469208211 )  ( -4 , 63.6044362292 )  ( -2 , 74.3468795356 )  ( 0 , 83.1386861314 )  ( 2 , 86.604870956 )  ( 4 , 89.5007342144 )  ( 6 , 89.9097472924 )  ( 8 , 90.5139765555 )  ( 10 , 89.9817850638 )  ( 12 , 89.6976281007 )  ( 14 , 89.333825429 )  ( 16 , 89.5384615385 )  ( 18 , 89.7749774977 ) 
};
\addlegendentry{Depth:1, Ncells:128}
\addplot[mark=otimes, color=violet] coordinates { 
( -20 , 9.80786825252 )  ( -18 , 9.13714804723 )  ( -16 , 9.73227206946 )  ( -14 , 11.0068567304 )  ( -12 , 14.0747493163 )  ( -10 , 20.7215212369 )  ( -8 , 32.2616204299 )  ( -6 , 48.0571847507 )  ( -4 , 60.6099815157 )  ( -2 , 72.351233672 )  ( 0 , 79.8722627737 )  ( 2 , 82.6608505998 )  ( 4 , 84.4713656388 )  ( 6 , 85.2166064982 )  ( 8 , 85.572587917 )  ( 10 , 85.883424408 )  ( 12 , 85.2435270686 )  ( 14 , 85.1817678538 )  ( 16 , 84.6877828054 )  ( 18 , 85.3645364536 ) 
};
\addlegendentry{Depth:1, Ncells:256}
\end{axis}
\end{tikzpicture}
\caption{Classification accuracy of single layer amplitude-phase \ac{lstm} model for different cell size on RadioML dataset.}
\label{fig_depth_1}
\end{figure}

\subsection{Learned representations}

The inherent non-linearity and deep structures makes understanding the representations learned by \ac{lstm}s difficult. In order to obtain some good insights we use visualization techniques similar to the ones presented in \cite{vis_karpathy}. These visualizations can help to understand how \ac{lstm} cells behave for an input signal, for instance which cells gets activated at each time step and how long each gate remains open. Figures~\ref{fig_activations} and \ref{fig_gatesat} presents the gate activation and saturation of the trained two layer \ac{lstm} model for a QAM64 input signal with 18dB \ac{snr}. As explained in Section~\ref{lstm_primer} the gates of \ac{lstm} cells have sigmoid activation functions, giving an output value between 0 and 1. A gate is said to be left saturated if its activation is less than 0.1 and right saturated if the activation is greater than 0.9. The fraction of time for which the gate is in left or right saturated mode in the entire 128 samples time is plotted in Figure~\ref{fig_gatesat}. On the first layer, it can be noticed that all the three gates are confined close to the origin showing that they are not highly left or right saturated. The absence of right saturation in the first layer forget gates, confirms that the cells do not store information for long term. There are no cells in the first layer that function in purely feed-forward fashion, since their forget gates would show up as consistently left-saturated. The output gate plots in the first layer also show that there are no cells that are revealed or blocked to the hidden state. This is also visible in the activation plots of the first layer in Figure~\ref{fig_activations}. The activations are short when compared to the second layer and it can be noticed in Figure~\ref{fig_activations} that many cell activations follow the input amplitude and phase changes in the input waveform. The second layer stores much long term dependencies from the fine grained representations generated from the first layer. 

\subsection{Effect of cell size and layer depth}
\label{sec_hyperparams}

\begin{figure}[htb]
\begin{tikzpicture} \begin{axis}[legend pos=south east,
 width=\columnwidth,
 grid=both,
 ylabel=Classification Accuracy (\%),
  xlabel=SNR(dB),
 grid style={line width=.1pt, draw=gray!10},
 major grid style={line width=.2pt,draw=gray!50},
 minor tick num=5,
 xmin=-20,
 xmax=20,
 ymax=100,
 legend cell align={left},
]

\addplot[mark=square, color=black] coordinates { 
( -20 , 9.51509606587 )  ( -18 , 9.46412352407 )  ( -16 , 9.62373371925 )  ( -14 , 10.8444604836 )  ( -12 , 14.8040109389 )  ( -10 , 21.0769504176 )  ( -8 , 29.5241594709 )  ( -6 , 42.8885630499 )  ( -4 , 51.146025878 )  ( -2 , 56.3497822932 )  ( 0 , 57.8832116788 )  ( 2 , 58.524173028 )  ( 4 , 59.8751835536 )  ( 6 , 60.3429602888 )  ( 8 , 59.1704238052 )  ( 10 , 59.1985428051 )  ( 12 , 58.9715734202 )  ( 14 , 59.5866396014 )  ( 16 , 58.7330316742 )  ( 18 , 60.3780378038 ) 
};
\addlegendentry{Depth:2, Ncells:16}
\addplot[mark=triangle, color=blue] coordinates { 
( -20 , 9.53339432754 )  ( -18 , 9.22797456857 )  ( -16 , 9.69609261939 )  ( -14 , 10.7722843739 )  ( -12 , 14.2935278031 )  ( -10 , 22.3742669273 )  ( -8 , 33.731398126 )  ( -6 , 49.3768328446 )  ( -4 , 63.1423290203 )  ( -2 , 71.5711175617 )  ( 0 , 78.0474452555 )  ( 2 , 81.0432569975 )  ( 4 , 83.2232011747 )  ( 6 , 83.6281588448 )  ( 8 , 84.869251578 )  ( 10 , 84.2258652095 )  ( 12 , 82.9802643491 )  ( 14 , 84.203727625 )  ( 16 , 83.257918552 )  ( 18 , 83.6543654365 ) 
};
\addlegendentry{Depth:2, Ncells:32}
\addplot[mark=o, color=red] coordinates { 
( -20 , 9.38700823422 )  ( -18 , 9.11898274296 )  ( -16 , 9.87698986975 )  ( -14 , 11.2955611693 )  ( -12 , 14.2388331814 )  ( -10 , 21.8233516972 )  ( -8 , 34.0804703289 )  ( -6 , 50.9530791789 )  ( -4 , 64.7134935305 )  ( -2 , 74.7641509434 )  ( 0 , 82.1897810219 )  ( 2 , 85.1872046529 )  ( 4 , 86.7657856094 )  ( 6 , 87.4007220217 )  ( 8 , 87.7186654644 )  ( 10 , 86.8670309654 )  ( 12 , 87.1808799565 )  ( 14 , 86.6949621701 )  ( 16 , 86.9321266968 )  ( 18 , 87.2907290729 ) 
};
\addlegendentry{Depth:2, Ncells:64}
\addplot[mark=diamond, color=brown] coordinates {
( -20 , 9.64318389753 )  ( -18 , 9.40962761126 )  ( -16 , 10.003617945 )  ( -14 , 12.3962468423 )  ( -12 , 15.9890610757 )  ( -10 , 23.1384396659 )  ( -8 , 34.3927980893 )  ( -6 , 48.4237536657 )  ( -4 , 63.4011090573 )  ( -2 , 76.2880986938 )  ( 0 , 84.6532846715 )  ( 2 , 88.9312977099 )  ( 4 , 89.7577092511 )  ( 6 , 90.4151624549 )  ( 8 , 90.9287646528 )  ( 10 , 90.2732240437 )  ( 12 , 90.3856599674 )  ( 14 , 90.3118656579 )  ( 16 , 90.3167420814 )  ( 18 , 90.5850585059 ) 
};
\addlegendentry{Depth:2, Ncells:128}
\addplot[mark=otimes, color=violet] coordinates { 
( -20 , 9.88106129918 )  ( -18 , 9.28247048138 )  ( -16 , 10.5463096961 )  ( -14 , 11.6564417178 )  ( -12 , 15.7338195077 )  ( -10 , 22.6763817309 )  ( -8 , 33.6762814624 )  ( -6 , 49.7983870968 )  ( -4 , 65.1756007394 )  ( -2 , 76.8142235123 )  ( 0 , 84.9087591241 )  ( 2 , 87.4772809887 )  ( 4 , 89.1703377386 )  ( 6 , 90.1624548736 )  ( 8 , 90.1352569883 )  ( 10 , 90.1639344262 )  ( 12 , 90.0054318305 )  ( 14 , 89.3707326075 )  ( 16 , 89.7737556561 )  ( 18 , 89.8469846985 ) 
};
\addlegendentry{Depth:2, Ncells:256}
\end{axis}
\end{tikzpicture}
\caption{Classification accuracy of two layer amplitude-phase \ac{lstm} model for different cell sizes on RadioML dataset.}
\label{fig_depth_2}
\end{figure}
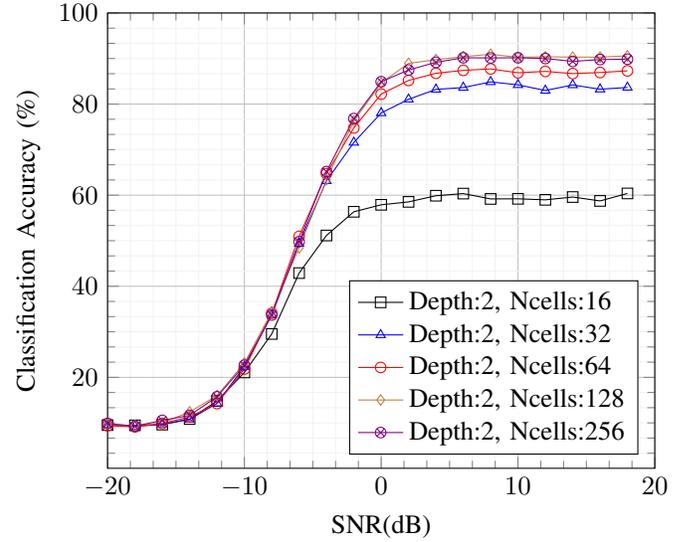

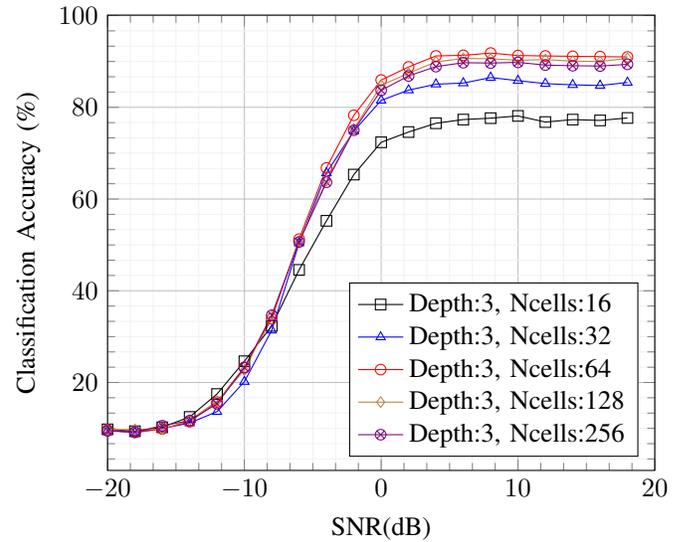
\begin{figure}[htb]
\begin{tikzpicture} \begin{axis}[legend pos=south east,
 width=\columnwidth,
 grid=both,
 ylabel=Classification Accuracy (\%),
  xlabel=SNR(dB),
 grid style={line width=.1pt, draw=gray!10},
 major grid style={line width=.2pt,draw=gray!50},
 minor tick num=5,
 xmin=-20,
 xmax=20,
 legend cell align={left},
]

\addplot[mark=square, color=black] coordinates { 
( -20 , 9.78956999085 )  ( -18 , 9.37329700272 )  ( -16 , 10.2387843705 )  ( -14 , 12.4864669794 )  ( -12 , 17.484047402 )  ( -10 , 24.631242225 )  ( -8 , 32.3351093147 )  ( -6 , 44.5564516129 )  ( -4 , 55.2495378928 )  ( -2 , 65.3301886792 )  ( 0 , 72.3722627737 )  ( 2 , 74.6092330062 )  ( 4 , 76.5234948605 )  ( 6 , 77.3285198556 )  ( 8 , 77.6375112714 )  ( 10 , 78.1238615665 )  ( 12 , 76.7879775484 )  ( 14 , 77.3205388448 )  ( 16 , 77.1402714932 )  ( 18 , 77.6777677768 ) 
};
\addlegendentry{Depth:3, Ncells:16}
\addplot[mark=triangle, color=blue] coordinates { 
( -20 , 9.47849954254 )  ( -18 , 9.11898274296 )  ( -16 , 9.98552821997 )  ( -14 , 11.1692529773 )  ( -12 , 13.6371923428 )  ( -10 , 20.1883774658 )  ( -8 , 31.3797538122 )  ( -6 , 50.4215542522 )  ( -4 , 65.6931608133 )  ( -2 , 74.8367198839 )  ( 0 , 81.5145985401 )  ( 2 , 83.7513631407 )  ( 4 , 85.0403817915 )  ( 6 , 85.2888086643 )  ( 8 , 86.4923354373 )  ( 10 , 85.810564663 )  ( 12 , 85.1711026616 )  ( 14 , 84.9049640155 )  ( 16 , 84.7601809955 )  ( 18 , 85.4185418542 ) 
};
\addlegendentry{Depth:3, Ncells:32}
\addplot[mark=o, color=red] coordinates { 
( -20 , 9.66148215919 )  ( -18 , 9.10081743869 )  ( -16 , 9.87698986975 )  ( -14 , 11.4579574161 )  ( -12 , 15.7338195077 )  ( -10 , 23.3339257153 )  ( -8 , 33.8967481168 )  ( -6 , 51.2646627566 )  ( -4 , 66.7837338262 )  ( -2 , 78.2656023222 )  ( 0 , 85.9489051095 )  ( 2 , 88.8040712468 )  ( 4 , 91.2077826725 )  ( 6 , 91.3357400722 )  ( 8 , 91.830477908 )  ( 10 , 91.2932604736 )  ( 12 , 91.2004345464 )  ( 14 , 91.1238235837 )  ( 16 , 91.0588235294 )  ( 18 , 90.9810981098 ) 
};
\addlegendentry{Depth:3, Ncells:64}
\addplot[mark=diamond, color=brown] coordinates { 
( -20 , 9.69807868253 )  ( -18 , 9.77293369664 )  ( -16 , 9.80463096961 )  ( -14 , 11.7827499098 )  ( -12 , 15.934366454 )  ( -10 , 22.69415319 )  ( -8 , 34.5948925225 )  ( -6 , 50.7881231672 )  ( -4 , 64.0480591497 )  ( -2 , 75.2539912917 )  ( 0 , 84.8175182482 )  ( 2 , 87.4045801527 )  ( 4 , 89.922907489 )  ( 6 , 90.6498194946 )  ( 8 , 90.5861136159 )  ( 10 , 90.2732240437 )  ( 12 , 90.4037660692 )  ( 14 , 90.071968998 )  ( 16 , 89.9909502262 )  ( 18 , 90.6750675068 )
};
\addlegendentry{Depth:3, Ncells:128}
\addplot[mark=otimes, color=violet] coordinates { 
( -20 , 9.46020128088 )  ( -18 , 9.22797456857 )  ( -16 , 10.5824891462 )  ( -14 , 11.6925297726 )  ( -12 , 15.3509571559 )  ( -10 , 23.2272969611 )  ( -8 , 34.6683814073 )  ( -6 , 50.6414956012 )  ( -4 , 63.6598890943 )  ( -2 , 75.0 )  ( 0 , 83.704379562 )  ( 2 , 86.8411486732 )  ( 4 , 88.8766519824 )  ( 6 , 89.7292418773 )  ( 8 , 89.6483318305 )  ( 10 , 89.7996357013 )  ( 12 , 89.2087633533 )  ( 14 , 89.0754751799 )  ( 16 , 88.9954751131 )  ( 18 , 89.3789378938 )
};
\addlegendentry{Depth:3, Ncells:256}
\end{axis}
\end{tikzpicture}
\caption{Classification accuracy of three layer amplitude-phase \ac{lstm} model for different cell sizes on RadioML dataset.}
\label{fig_depth_3}
\end{figure}

A comprehensive study is also performed to understand the effect of the number cells and layer depth on the model performance. The number of \ac{lstm} cells and layer depth are varied from 16 to 256 and  1 to 3 respectively. The models are trained on RadioML dataset on all \ac{snr}s. The accuracy levels for various layer depths are presented in Figures~\ref{fig_depth_1}, \ref{fig_depth_2} and \ref{fig_depth_3}. An initial analysis clearly shows that the model accuracy increases with increasing layer depth for mostly all cell sizes. It can be also noted that as the depth of the model increases, increasing the number of cells doesn't give much performance improvements. For instance, at depths 2 and 3 increasing the cell numbers from 128 to 256 doesn't provide any performance improvements.

\section{Resource Friendly Models}
\label{lmodels}
As mentioned earlier in the introduction, it is quite difficult to deploy \ac{soa} \ac{amc} algorithms on low-end distributed sensors such as the ones in Electrosense. We extend our study in two directions to reduce the resource requirements in terms of data transfer rate to the cloud, data storage and computational power. First, a study is conducted to understand to what extent technology classification based on \ac{amc} can be done using averaged magnitude \ac{fft} data, as the \ac{psd} pipeline being the default enabled one in the Electrosense sensors with medium data transfer costs. As the sensors are sequentially scanning the spectrum, they are capable of generating magnitude spectrum information for wideband signals which is an added advantage. Second, the performance of quantized versions of the proposed deep learning models are analyzed which can reduce the computational cost of the models enabling deployment of these models on the sensors itself. The averaged magnitude \ac{fft} signal sent by the sensor, the selected dataset for testing the model, averaged magnitude \ac{fft} classification model, other quantized models and the classification results are detailed in the following subsections.

\subsection{Received averaged magnitude \ac{fft} signal}
Electrosense sensors scan the wireless spectrum sequentially. The sensor samples the wireless spectrum at a fixed sampling rate $N = 2.4~MS/s$ tuned to a particular centre frequency $f_x$.  As the sensor's sampling rate is limited, a wideband signal's magnitude spectrum can be received only by sequential scans to cover the entire bandwidth as given in equation~\ref{eq_avg}.
\begin{equation}
\begin{split}
R(f)&=\dfrac{1}{M}\sum_{m=0}^{M}|FFT_m(e^{-j2\pi f_1(t_0+mD_t)}s(t_0+mD_t)*\\ &c(t_0+mD_t)+n(t_0+mD_t))|\\ 
&||\dfrac{1}{M}\sum_{m=0}^{M}|FFT_m(e^{-j2\pi f_2(t_1+mD_t)}s(t_1+mD_t)*\\
&c(t_1+mD_t)+n(t_1+mD_t))|\\
&||\ldots
\end{split}
\label{eq_avg}
\end{equation}

In equation~\ref{eq_avg} $||$ represents the concatenation operation where the full bandwidth of the signal of interest is captured by a sequentially scanning sensor sampling at a lower sampling rate, similar to the Electrosense dataset mentioned in the following subsection. The averaged magnitude \ac{fft} signal at centre frequencies $f_i$, where $f_i \in (50~MHz, 6~GHz)$ based on the sensor sampling rate and frequency range, are sent to the cloud where they are concatenated together. In equation~\ref{eq_avg}, $M$ represents the magnitude-\ac{fft} averaging factor and $t_x$ the sequential sampling time. For instance, $t_n = t_{n-1}+T$, where $T=MD_t$ is the amount of time spent at a particular centre frequency and $D_t$ being the time for collecting \textit{fft\_size} samples for a single FFT input.

\subsection{Electrosense dataset}
\label{electrosense_dataset}
Six commercially deployed technologies are selected to validate the classification accuracy using averaged magnitude \ac{fft} data as given in Table~\ref{table_elec_dataset}. Over-the-air data from multiple Electrosense sensors are retrieved through the Electrosense API\footnote{\label{noteapi}https://electrosense.org/open-api-spec.html} with a spectral resolution of 10~kHz and time resolution of 60 seconds. The data is collected from sensors with omni-directional antennas which are deployed indoors. The sensors follow sequential scanning of the spectrum with an \ac{fft} size set to 256 giving a frequency resolution close to 10~kHz. With a \ac{fft} size of 256 and sensor ADC bit-width of 8, we get an effective bitwidth of 12 resulting in a theoretical dynamic range of 74dB. Practical dynamic range depends on the ADC frontend stages and the noise level, which may vary between 60 to 65dB. Five \ac{fft} vectors are averaged for reducing the thermal noise of the receiver. Some of the selected technologies such as LTE and DVB have an effective bandwidth which is higher than the sampling bandwidth of the of the low-end \ac{sdr}. As the sensor is sequentially scanning, full spectrum shapes of these wideband signals are obtained by combining \ac{fft} outputs of these sequential scans. The entire data is split into two, one half for training and the other half for testing the model. 

\begin{table}[!t]
\begin{center}
\begin{tabular}{|l|l|}
	\hline
    Technology     & WFM, TETRA, DVB,\\  
                   & RADAR, LTE, GSM \\
    								 
    \hline
    Time resolution & 60s\\
    \hline
    Frequency resolution & 10~kHz\\
  	\hline
    Sensor sampling rate &  2.4 MS/s \\
  	\hline
    Scanning strategy &  Sequential \\
  	\hline
    FFT averaging factor  & 5 \\
    \hline 
    Sensor location and antenna type & Indoor, Omni-directional\\    
    \hline
  	Number of training samples &   3100 vectors\\
  	\hline
  	Number of test samples &  3100 vectors\\
    \hline
\end{tabular}
\end{center}
\caption{Averaged magnitude \ac{fft} dataset parameters.}
\label{table_elec_dataset}
\end{table}

\subsection{Averaged Magnitude \ac{fft} model}
Sequentially sensed frequency spectrum data from the sensors contain signals of different bandwidth. The model should be able to process this variable length data 
and classify them to proper groups. We use the same \ac{lstm} model used for classifying complex input data as shown in Figure~\ref{fig_iq_lstm_model}. 
The averaged magnitude \ac{fft} signal is fed to the model as a sequence as presented in Figure~\ref{fig_iq_lstm_model}. The same \ac{lstm} model is chosen as it can handle variable length input signals and is also good at learning long term dependencies. The final output of the \ac{lstm} model after feeding $n$ frequency bins 
is given as input to the softmax layer through a fully connected linear layer. The softmax layer outputs the probability $P(y=l|a;\theta)$ for $l \in \{0,1,..,5\}$ where $a$ denotes the input averaged \ac{fft} bins, $\theta$ the model parameters and $l$ the enumerated label for one of the six technologies as listed in Table~\ref{table_elec_dataset}.

\begin{figure}[!t]
\centering
\includegraphics[width=1.1\columnwidth]{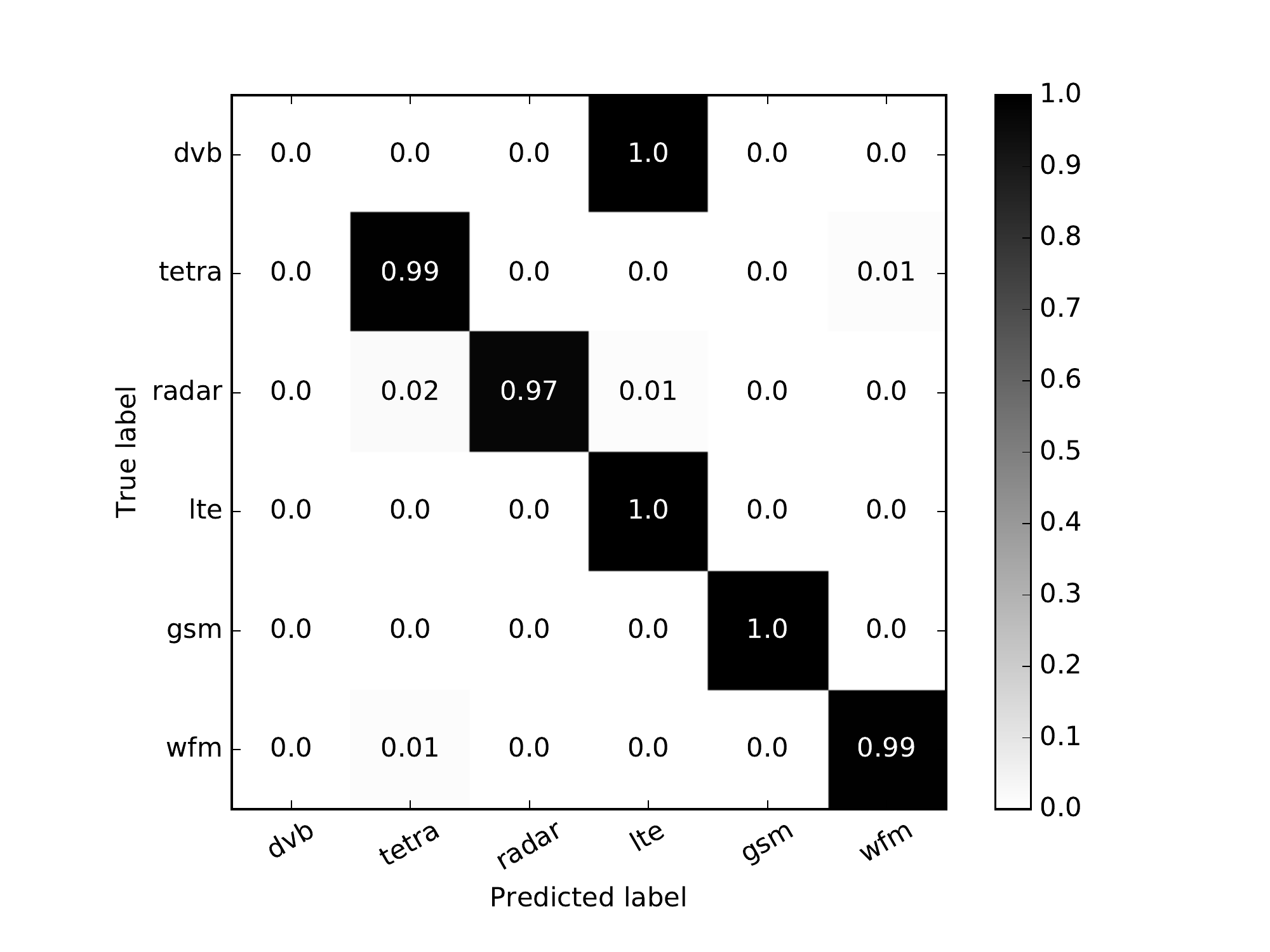}
\caption{Confusion matrix for technology classification on Electrosense dataset.} 
\label{fig_confmat_tech}
\end{figure}

\begin{table}[!t]
\begin{center}
\begin{tabular}{|c|l|l|l|l|}
  \hline
  \multirow{2}{*}{Layer-depth} 
      & \multicolumn{4}{c|}{Number of cells} \\
      \cline{2-5}
  & 16 & 32 & 64 & 128 \\  \hline
  1 & 70.02\% & 74.39\% & 79.38\% & 81.05\%\\      \hline
  2 & 72.69\% & 75.93\% & 80.92\% & 81.68\%\\      \hline
\end{tabular}
\end{center}
\caption{Classification accuracy on Electrosense dataset for varying layer depths and cell numbers.}
\label{table_magfft_classif}
\end{table}

\begin{figure}[!t]
\centering
\includegraphics[width=1\columnwidth]{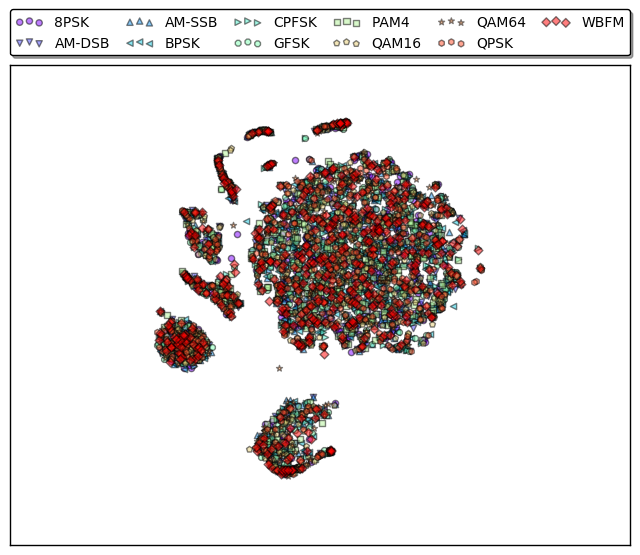}
\caption{t-SNE plot for magnitude FFT output on radioml dataset.} 
\label{fig_tsne}
\end{figure}

\subsection{Classification results}
An initial study is conducted to understand the technology classification accuracy of the averaged magnitude \ac{fft} model when compared to full IQ information. On the Electrosense dataset (Section \ref{electrosense_dataset}) the proposed model achieves a classification accuracy of 80\%. The confusion matrix for the same is shown in Figure~\ref{fig_confmat_tech}.  From the confusion matrix it is clear that there is a large confusion between LTE and DVB. This is expected as the power spectra of both DVB and LTE looks very similar as both of them are based on \ac{ofdm}. As multiple technologies might share the same modulation types, the assumption that a modulation classifier can be used for technology classification is not always valid with the current deployed technologies. We also investigated the effect of number of \ac{lstm} cells and layer depth on this dataset whose results are summarized in Table~\ref{table_magfft_classif}. Increasing the layer depth did not contribute significantly to the classification accuracy as there might be no more low level abstract features to be learned from the magnitude spectrum information. Furthermore, there are a large number of modulation schemes which exhibit the same power spectral density making averaged spectrum a sub-optimal feature for classification. For instance, the power spectral densities of different modulations schemes such as 8PSK, QAM16 and QPSK are identical once passed through a pulse shaping filter with a fixed roll-off factor. This can be theoretically shown and easily verified with manual inspection \cite{digital_psd}. 

To further validate the argument, magnitude-\ac{fft} is calculated on the same RadioML dataset which was used for testing the performance of the amplitude-phase model. As the RadioML dataset consists of modulations with same bandwidths passed through the same pulse shaping filter, their magnitude-\ac{fft}s looks identical giving very low classification accuracy of only 19\% for all 11 modulations even at high \ac{snr}s. To get a better understanding of the generated magnitude-\ac{fft} dataset, a visualization of a subset of the data in two dimensions is provided. For reducing the dimensionality of the data to $2$ and for the ease of plotting, the t-SNE technique \cite{tsne} is used. A small subset of the radioML dataset of 5000 vectors, containing 128 point magnitude FFT of all generated modulation schemes with varying SNR ranges from +20dB to +20dB are fed to the t-SNE algorithm. t-SNE is useful for a preliminary analysis to check whether classes are separable in some linear or nonlinear representation. The representation generated by t-SNE on the data subset is presented in Figure~\ref{fig_tsne}. It can be seen that the representation overlap is very high and t-SNE could not generate any meaningful clustering as the phase information is completely lost when computing magnitude-\ac{fft}, leaving identical magnitude spectrum for many modulation schemes. The obvious solution is to switch to \ac{iq} pipeline and deploy optimized versions of complex input signal models on the sensors itself, thus reducing the uplink data transfer rate. This is further investigated in the following subsection.

\subsection{Quantized models}
\begin{figure}[htb]
\begin{tikzpicture} \begin{axis}[legend pos=south east,
 width=\columnwidth,
 grid=both,
 ylabel=Classification Accuracy (\%),
  xlabel=SNR(dB),
 grid style={line width=.1pt, draw=gray!10},
 major grid style={line width=.2pt,draw=gray!50},
 minor tick num=5,
 xmin=-20,
 xmax=20,
 legend cell align={left},
]

\addplot[mark=square, color=black] coordinates { 
(-20, 09.807868252516011) (-18, 09.282470481380563) (-16, 09.659913169319827) (-14, 09.115341032118368) (-12, 15.332725615314494) (-10, 23.049582370712635) (-8, 37.6630534631637) (-6, 53.40909090909091) (-4, 63.99260628465804) (-2, 71.89767779390421) (0, 75.98540145985402) (2, 76.20865139949109) (4, 78.1754772393539) (6, 78.23104693140794) (8, 78.53922452660054) (10, 79.67213114754098) (12, 78.47184501176897) (14, 78.2432183059605) (16, 77.50226244343892) (18, 78.36183618361836) };
\addlegendentry{CNN: full precision}
\addplot[mark=otimes, color=purple] coordinates { 
( -20 , 9.95425434584 )  ( -18 , 9.08265213442 )  ( -16 , 9.69609261939 )  ( -14 , 10.9346806207 )  ( -12 , 12.7073837739 )  ( -10 , 16.6696285765 )  ( -8 , 22.9836487231 )  ( -6 , 32.1114369501 )  ( -4 , 41.8669131238 )  ( -2 , 52.3222060958 )  ( 0 , 61.9160583942 )  ( 2 , 65.2308251545 )  ( 4 , 66.3362701909 )  ( 6 , 67.8700361011 )  ( 8 , 67.6825969342 )  ( 10 , 68.306010929 )  ( 12 , 67.0650009053 )  ( 14 , 67.7062188596 )  ( 16 , 67.4027149321 )  ( 18 , 68.0648064806 ) };
\addlegendentry{CNN: binary}
\addplot[mark=o, color=blue] coordinates {
(-20, 09.387008234217749) (-18, 08.991825613079019) (-16, 09.334298118668596) (-14, 10.285095633345363) (-12, 12.178669097538743) (-10, 35.64954682779456) (-8,  52.36083042439831) (-6,  67.46700879765396) (-4,  76.7097966728281) (-2,  82.89187227866474) (0, 88.48540145985402) (2, 90.27626317702654) (4, 92.0704845814978) (6, 91.85920577617328) (8, 92.19116321009919) (10, 91.74863387978142) (12, 91.56255658156799) (14, 91.58516331426463) (16, 91.13122171945701) (18, 91.82718271827183)
};
\addlegendentry{LSTM: full precision}
\addplot[mark=triangle, color=red] coordinates {
(-20, 09.307868252516011) (-18, 09.132470481380563) (-16, 09.479913169319827) (-14, 09.115341032118368) (-12, 10.847725615314494)(-10,15.150416817687568) (-8, 30.853792598021251) (-6, 47.477477477477475) (-4, 56.218637992831544) (-2, 63.014450338394001) (0,  69.974508375819378) (2,  73.602656336469285) (4,  76.295479603087102) (6,  79.15686629274471) (8,  78.14148026913984) (10, 78.852849551035364) (12, 79.740828618361015) (14, 79.506573023590854) (16, 79.46428571428571) (18, 79.637277787753635)
};
\addlegendentry{LSTM: TW\_4BA}
\addplot[mark=diamond, color=brown] coordinates {
(-20, 09.368709972552608) (-18, 09.264305177111716) (-16, 09.280028943560058) (-14, 09.545290508841574) (-12, 11.777575205104832) (-10, 16.527456904211835) (-8,  28.036009553555025) (-6,  42.906891495601174) (-4,  57.7634011090573) (-2,  69.97460087082729) (0,   75.62043795620438) (2,   80.57070156306797) (4,   80.9287812041116) (6,   83.44765342960289) (8,   84.56266907123535) (10,  84.24408014571949) (12,  83.86746333514394) (14,  84.01919173279203) (16,  84.61538461538461) (18,  84.98649864986498)
};
\addlegendentry{LSTM: TW\_FA}
\end{axis}
\end{tikzpicture}
\caption{Classification accuracy of two layer quantized models on RadioML dataset.}
\label{fig_quant_acc}
\end{figure}
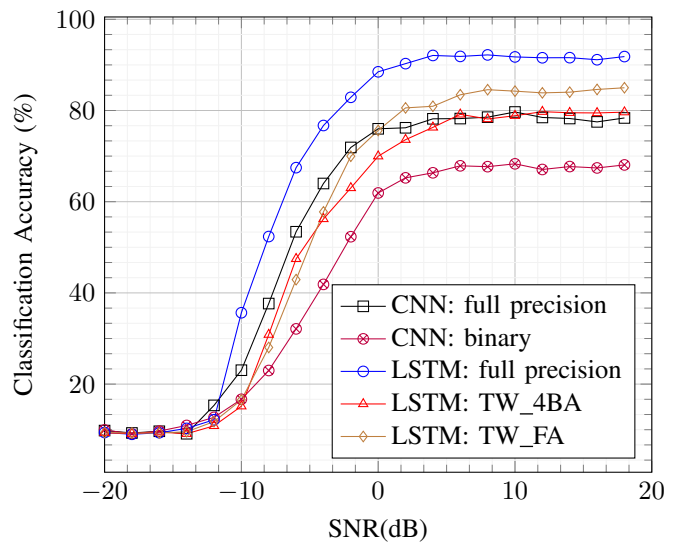

Deep learning models are processor intensive which makes them quite difficult to be deployed on low-end sensor networks. As mentioned in the introduction, transferring pure \ac{iq} data to the backend server is not a scalable solution. In addition, our results indicate that some signals require \ac{iq} information for correct classification. To enable low-end sensor deployment, a feasibility study is conducted by quantizing the weights and activations of the newly proposed as well as baseline neural network models. Binarized networks can exceptionally reduce the required memory size for holding intermediate results and replace most of the arithmetic operations with bitwise operations \cite{quant_neural}. For instance, when compared to a full precision network with 32 bits weights and activation, a binarized network only needs 32 times smaller memory resulting in a reduced required memory size and memory access cost. In \cite{quant_neural} the authors had already noticed that binarizing \ac{lstm}s results in very poor accuracy. We confirm the same observation with our \ac{lstm} models too. However, models with binarized \ac{cnn}s have been reported to provide accuracy close to their full precision variants. To validate this, the performance of the binarized baseline \ac{cnn} model is also investigated on the radioML modulation dataset. Furthermore, by allowing more quantization levels on the \ac{lstm} models a higher accuracy can be achieved while still reducing the computational cost. Two quantized \ac{lstm} model variants are tested, one with ternary weights (-1, 0, +1) and full precision activation (TW\_FA) and the other with ternary weights and four bits activation (TW\_4BA). The accuracy results of these models are summarized in Figure~\ref{fig_quant_acc}. Results show that \ac{lstm} models with ternary weights and 4bit activation can provide close to 80\% accuracy reducing the very high computational power required for full precision models. Binary \ac{cnn} models also provided an accuracy level 10\% below the full precision variants. We believe the classification accuracy can be further improved by proper hyper-parameter tuning and longer training.

\begin{table}[!t]
\begin{center}
\begin{tabular}{|l|l|l|l|}
	\hline
    Model     & Weights count & MUL/ & Memory\\
    & & XNOR count & \\
    \hline
    CNN: full precision & 5451096 & 5451096 & 174.44~MB\\
    \hline
    CNN: binary & 5451096 & 5451096 & 5.45~MB\\
    \hline
    LSTM: full precision & 200075 & 262144 & 6.4~MB\\
    \hline
    LSTM: TW\_4BA & 200075 & 262144 & 400.15~KB\\
    \hline
    LSTM: TW\_FA & 200075 & 262144 &400.15~KB\\
    \hline
\end{tabular}
\end{center}
\caption{Memory size and number of multiplications for the discussed models (excluding nonlinear activations).}
\label{table_quant_perf}
\end{table}

\begin{table}[!t]
\begin{center}
\resizebox{\columnwidth}{!}{
\begin{tabular}{|l|l|l|l|l|}
  \hline
  \multirow{2}{*}{Platform} 
      & \multicolumn{4}{c|}{Model} \\
      \cline{2-5}
  & CNN-FP & CNN-8b & LSTM-FP & LSTM-8b \\  \hline
  Nvidia GeForce GTX 1060 & 15239  & 85  & 12874  & 76 \\      \hline
  Nvidia Tegra TX2 & 2342 & 205 & 2785 & 127 \\      \hline
  Intel i7-3770 
  & 1039  & 85 & 1035 & 81 \\      \hline
  Jetson board (ARMv8) & 264 & 217 &  275 & 164\\      \hline
  Raspberry pi-2 (ARMv7) & 36 & 30 & 24 & 21\\      \hline
\end{tabular}
}
\end{center}
\caption{Inference engine performance in number of classification per second on different platforms.}
\label{table_perf_comp}
\end{table}

The theoretical memory requirements for the trained weights along with number of multiplications required for the entire model, excluding activations, are summarized in Table~\ref{table_quant_perf}. A binarized neural network can drastically reduce the processing power requirements of the model. For instance, in a binarized network all weights and activation are either -1 or +1, replacing all multiply operations by XNORs. The multipy-accumulate, which is the core operation in neural networks, can be replaced by 1-bit XNOR-count operation \cite{quant_neural}. Convolutions also comprises of multiply and accumulate operations which can also be replaced by its binarized variants. Thus the baseline \ac{cnn} model can provide very good performance improvements on the general purpose ARM based Electrosense sensors. For the CNN models the convolutional layer output numbers are high, as we are not using any pooling layers, which accounts for the larger memory size in the succeeding dense layers. We would like to emphasize the fact that the given memory sizes are for the entire model and the weights that should be hold in the memory might vary based on practical implementations. 

As binarized \ac{lstm} models did not provide good accuracy, we are forced to use 4-bit quantized variants of the same. Even though the performance improvements are not that extreme similar to binarized models, quantized \ac{lstm}s can also reduce the resource consumption. First of all, as no large dynamic range is required all the 4-bit multiply-accumulate operations can be implemented in fixed point arithmetic, which is much more faster in ARM CPUs when compared to their floating point versions. Secondly, routines can also be implemented to reduce the space requirements to hold intermediate results and the activations can be implemented as look-up tables. We would also like to emphasize the fact that on a special purpose hardware, such as FPGAs, quantized models can obviously reduce the space used  and power-consumption as the multiply-accumulate units have smaller bit-widths.

Most of the machine learning frameworks such as Tensorflow have started supporting quantized models for low-end processor deployment. The quantized kernels are under active development at the time of writing this paper which currently only supports a minimum quantization of 8 bit. Quantized kernels for all operations in all platforms are not available in these libraries resulting in low performance than expected. The full \ac{iq} information model classification performances on various platforms such as Nvidia GPUs, Intel and ARM processors for full precision and 8bit precision models are summarized in Table~\ref{table_perf_comp}. To avoid implementation mismatches across various models and platforms all the comparisons are done using quantization tools provided by Google's Tensorflow which is currently under active development. These tools allow to freeze and compress a trained model to a single file and then test it on various platforms easily. These values in the table are performance indicators in number of classifications per second for 128 sample length vectors. It can be noticed that the quantized models perform very bad on GPUs and Intel PCs due to lack of support. The quantized models currently provide performance very close to floating point variants on ARM processors. Quantized kernel support is improving for ARM processors due to increased demand for deploying these models on mobile devices. The aforementioned advantages of quantized models is expected to be available in the near future through these standard libraries.

\section{Conclusion and Future work}
\label{conclusion}
Wireless spectrum monitoring and signal classification over frequency, time and space dimensions is still an active research problem. In this paper we proposed a new \ac{lstm} based model which can classify signals with time domain amplitude and phase as input. \ac{soa} results on high \ac{snr}s (0 to 20dB) is achieved without using complex \ac{cnn}-\ac{lstm} models as mentioned in \cite{baseline}. Being a recurrent model we showed that the model can handle variable length inputs thus can capture sample rate variations effectively.

Though neural networks are good at function approximation, our experiments emphasize the fact that \emph{data preprocessing} and \emph{proper representation} are equally important. This claim is substantiated from our experiments with the \ac{lstm} model where the model gave poor results when fed with time domain \ac{iq} samples while it gave accuracies close to 90\% for high \ac{snr}s when provided with time domain amplitude and phase information. As shown by various \ac{soa} models in speech and image domains, performance improvements are seen with increasing layer depth which saturates after a few levels. In addition, we showed that basic technology classification is achievable by only using averaged magnitude \ac{fft} information over a distributed set of sensors complying with the uplink bandwidth resource constraints. Furthermore, experiments showed that quantized \ac{lstm} models can achieve good classification results thus reducing the processing power requirements at the cost of 10\% accuracy loss. This allows the deployment of these models on low cost sensors networks such as Electrosense enabling a wide area deployment. It is also remarkable that these deep learning models can classify signals with a fewer number of samples when compared to the expert feature variants, such as cyclic frequency estimators, enabling faster classification. Furthermore, deep learning allows for incremental learning, thus it would not be required to retrain the entire network from scratch for the new wireless non-idealities like antenna patterns and sensitivity. In addition, dedicated hardware is gaining popularity to reduce the deep learning model's energy and memory footprints which demands quantized versions of the models.

Although the \ac{lstm} models perform very well at high \ac{snr} conditions, CNN models seems to provide an additional 5-10\% accuracy on the low \ac{snr} conditions (SNRs below -2dB) as shown in \cite{baseline}. Even though we are not able to replicate the results in \cite{baseline} (
because of hyperparameter tuning), it is reasonable to conclude that the learned filters in CNN for a fixed sample rate might give performance improvements for low SNR values. Furthermore, all the implemented code for the proposed models are made publicly available for reproducing and verifying the results presented in this paper and for supporting future research.

Low \ac{snr} performance of these \ac{soa} deep learning models could be further improved with the help of efficient blind denoising models. Models which can perform automated channel equalization and compensate receiver imperfections such as frequency offset can further improve the classification performance. The current radio deep learning models make use of layers which basically applies non-linearity after simple multiply-accumulate-add operations while it is well established in the research community that cyclic cumulants, which are generated by time-shifted multiplication and averaging of the input itself, performs well in the expert feature space. Deep learning models which can extract features similar to cyclic cumulants might improve the performance metrics.

The analysis would not be complete without emphasizing the limitations of the \ac{soa} deep learning models. First, the current complex models are tested on a dataset with normalized bandwidth parameters. Real life transmitted signals generally have varying symbol rates and bandwidths. Even though the variable length \ac{lstm} model is shown to be capable of adapting to these scenarios, further analysis is required to validate the claim. In future, models that can handle all possible spread spectrum modulations should be also tested. Second, the generalization capabilities of these models should be further investigated, in terms of performance of these models in unknown channel conditions and modulation parameters. Finally, most of the successful \ac{soa} models are supervised models which requires labeled training data. Labeling is a very tedious task which projects the importance of semi-supervised models for classification tasks. Published studies \cite{mlresource} on semi-supervised machine learning models for cellular network resource management validates the need for more semi-supervised models, which is also an active direction for future research. We believe deep learning models adapted to radio domain can help in understanding, analyzing and decision making in future complex radio environments.

\appendices




\ifCLASSOPTIONcaptionsoff
  \newpage
\fi



%
\bibliographystyle{IEEEtran}
\bibliography{sections/bibliography}

\begin{IEEEbiography}[{\includegraphics[width=1in,height=1.25in,clip,keepaspectratio]{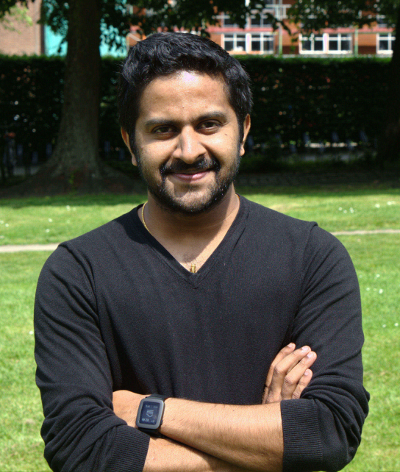}}]
    {Sreeraj Rajendran}
received his Masters degree in communication and signal processing from the Indian Institute of Technology, Bombay, in 2013. He is currently pursuing the PhD degree in the Department of Electrical Engineering, KU Leuven, Belgium. Before joining KU Leuven, he worked as a senior design engineer in the baseband team of Cadence and as an ASIC verification engineer in Wipro Technologies. His main research interests include machine learning algorithms for wireless and low power wireless sensor networks.
\end{IEEEbiography}

\begin{IEEEbiography}[{\includegraphics[width=1in,height=1.25in,clip,keepaspectratio]{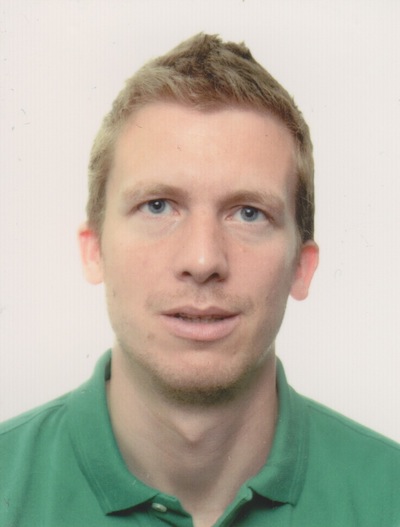}}]
    {Wannes Meert}
received his degrees of Master of Electrotechnical Engineering, Micro-electronics (2005), Master of Artificial Intelligence (2006) and Ph.D. in Computer Science (2011) from KU Leuven. He is currently research manager in the DTAI research group at KU Leuven. His work is focused on applying machine learning, artificial intelligence and anomaly detection technology to industrial application domains.
\end{IEEEbiography}

\begin{IEEEbiography}[{\includegraphics[width=1in,height=1.25in,clip,keepaspectratio]{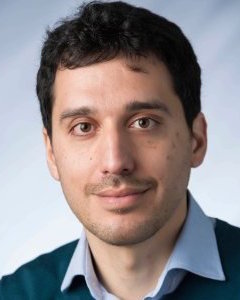}}]
    {Domenico Giustiniano}
is Research Associate Professor at IMDEA Networks Institute and leader of the Pervasive Wireless Systems Group. He was formerly a Senior Researcher and Lecturer at ETH Zurich and a Post-
Doctoral Researcher at Disney Research Zurich and at Telefonica Research Barcelona. He holds a Ph.D. from the University of Rome Tor Vergata (2008). He devotes most of his current research to visible light communication, mobile indoor localization, and collaborative spectrum sensing systems. He is an author of more than 70 international papers, leader of the OpenVLC project
and co-founder of the non-profit Electrosense association.
\end{IEEEbiography}

\begin{IEEEbiography}[{\includegraphics[width=1in,height=1.25in,clip,keepaspectratio]{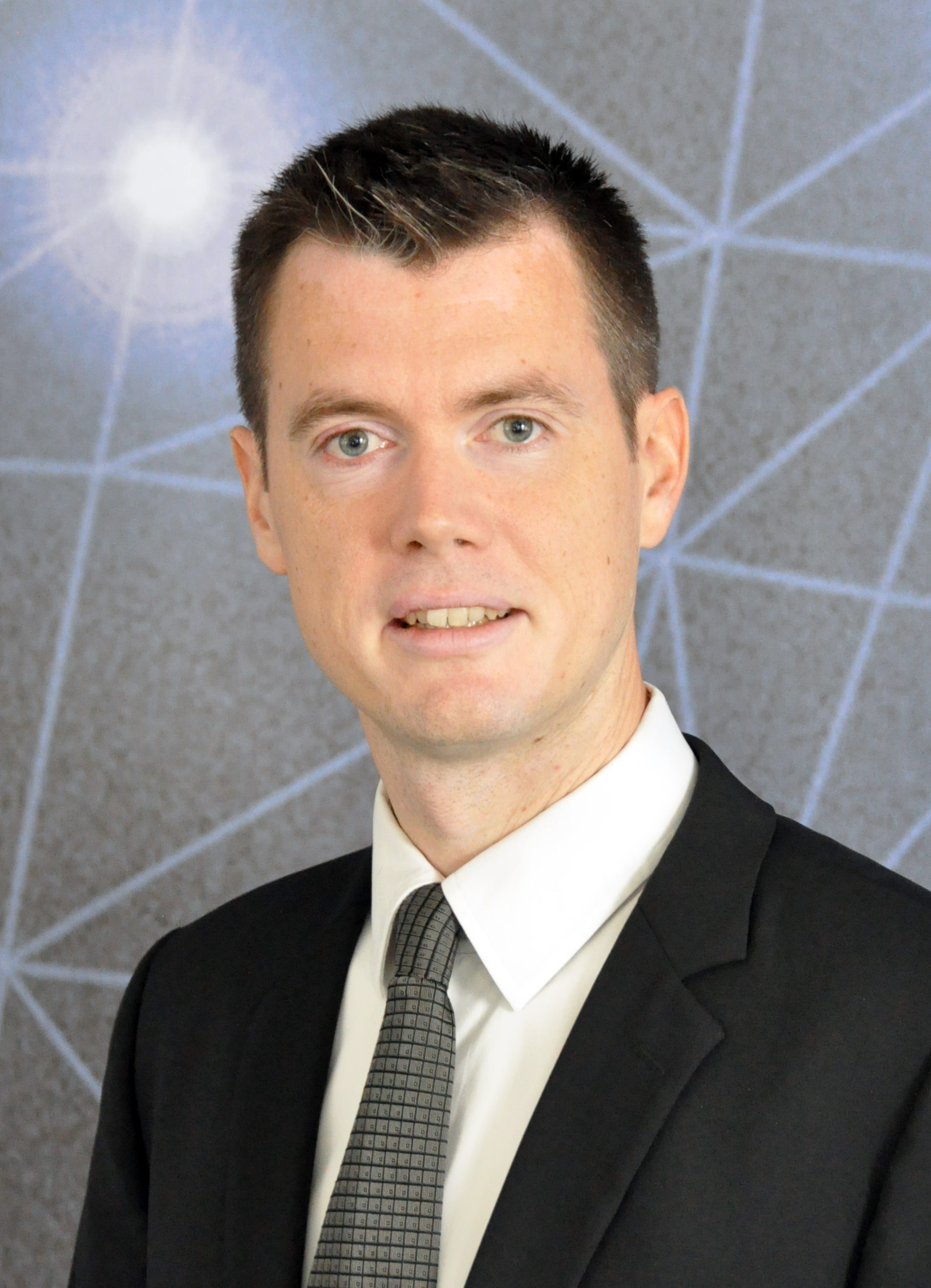}}]
    {Vincent Lenders}
is a research director at armasuisse where he leads the cyber and information sciences research of the Swiss Federal Department of Defense. He received the M.Sc. and Ph.D. degrees in electrical engineering from ETH Zurich. He was postdoctoral research fellow at Princeton University. Dr.Vincent Lenders is the cofounder and in the board of the OpenSky Network and Electrosense associations. His current research interests are in the fields of cyber security, information management, big data, and crowdsourcing.
\end{IEEEbiography}

\begin{IEEEbiography}[{\includegraphics[width=1in,height=1.25in,clip,keepaspectratio]{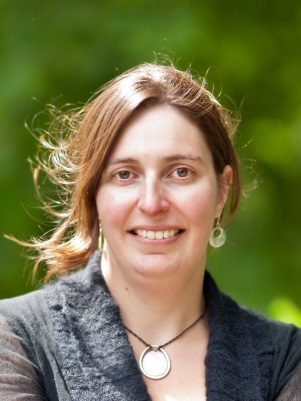}}]
    {Sofie Pollin}
obtained her PhD degree at KU Leuven with honors in 2006. From 2006-2008 she continued her research on wireless communication, energy-efficient networks, cross-layer design, coexistence and cognitive radio at UC Berkeley.  In November 2008 she returned to imec to become a principal scientist in the green radio team. Since 2012, she is tenure track assistant professor at the electrical engineering department at KU Leuven. Her research centers around Networked Systems that require networks that are ever more dense, heterogeneous, battery powered and spectrum constrained. Prof. Pollin is BAEF and Marie Curie fellow, and IEEE senior member. 
\end{IEEEbiography}
%








\end{document}